\documentclass[journal,twoside]{IEEEtran}
\usepackage{amsmath,amssymb,amsthm,bm,mathtools}
\usepackage{graphicx,booktabs,cite,url,xcolor}
\usepackage[hidelinks]{hyperref}
\hypersetup{pdftitle={Over-the-Scene Computation: Illumination-Programmed Functional Sensing of Unresolvable Populations},pdfauthor={Seyed Mohammad Azimi-Abarghouyi and Christopher G. Brinton}}
\newtheorem{theorem}{Theorem}
\newtheorem{proposition}{Proposition}
\newtheorem{lemma}{Lemma}
\newtheorem{corollary}{Corollary}

\newtheorem{assumption}{Assumption}

\newcommand{\C}{\mathbb{C}}
\newcommand{\R}{\mathbb{R}}
\newcommand{\E}{\mathbb{E}}
\newcommand{\Var}{\operatorname{Var}}
\newcommand{\Cov}{\operatorname{Cov}}
\newcommand{\tr}{\operatorname{tr}}
\newcommand{\supp}{\operatorname{supp}}
\newcommand{\rank}{\operatorname{rank}}
\newcommand{\sinc}{\operatorname{sinc}}
\newcommand{\Tsh}{\mathcal{T}}

\newcommand{\sS}{\mathcal{S}}

\newcommand{\sP}{\mathcal{P}}
\newcommand{\ip}[2]{\langle #1,#2\rangle}

\newcommand{\norm}[1]{\left\|#1\right\|}
\newcommand{\cc}{c}
\newcommand{\Tst}{T^{\star}}
\newcommand{\Fs}{\mathcal{F}}
\newcommand{\HS}{\mathrm{HS}}

\begin{document}

\title{Over-the-Scene Computation: Illumination-Programmed Functional Sensing of Unresolvable Populations}
\author{{Seyed Mohammad Azimi-Abarghouyi,~\IEEEmembership{Member,~IEEE}, and Christopher G. Brinton,~\IEEEmembership{Senior Member,~IEEE}}%
\thanks{S. M. Azimi-Abarghouyi is with the Department of Electronic Systems, Aalborg University, Copenhagen, Denmark (Email: smaa@es.aau.dk). C. G. Brinton is with the School of Electrical and Computer Engineering, Purdue University, West Lafayette, IN USA (Email: cgb@purdue.edu).}}
\IEEEaftertitletext{\vspace{-2\baselineskip}}
\maketitle

\begin{abstract}
Decisions about dense populations depend on how much will cross a boundary by a deadline and when the first member arrives. Conventional radar forms a range--Doppler map, whose cells add noise and whose detections merge unresolved members. We introduce over-the-scene computation, where illumination makes the scene compute such functionals without resolving members. With random echo phases, a correlator's mean output energy sums member reflectivities weighted by a kernel set by the waveform and template. The waveform's ambiguity function is this computation's transfer function, determining computable functionals and their noise cost. One readout aggregates members over a small delay--Doppler area but along a line segment growing with duration, longest for the chirp of matching slope among constant-envelope waveforms. Members predicted to share a range lie on such a line, so a chirp's mean beat spectrum is the predicted range profile at a look-ahead horizon set by its rate. For unbiased estimation, a Cram\'er--Rao bound over all equal-energy waveforms favors resolving members above unit signal-to-noise ratio (SNR) and grouping weaker ones into unit-SNR sets. At 0.65 members per cell, $-$10~dB member SNR, and one sweep, the chirp's bin sum inside the boundary comes within 16\% of this bound in root-mean-square error and, with 77 times fewer samples per horizon, outperforms every unbiased estimator from a coherent processing interval of equal dwell, bandwidth, and energy. No finite set of waveform--template probes determines an unconstrained Rayleigh-fluctuating population's first arrival, but bisection localizes it if first arrivals outweigh later arrivals' sidelobes.
\end{abstract}


\section{Introduction}\label{sec:intro}
\IEEEPARstart{F}{or} dense scatterer populations, such as birds, crowds, or blood cells, many decisions rest on functionals of the population rather than on its members, which may be too closely spaced or too weak to resolve. Whether to pause wind turbines or warn aircraft depends on how much of a migrating flock will cross a boundary within a given time and when its first member arrives. Current systems sense first and compute later (sense-then-compute): a coherent processing interval (CPI) forms a range--Doppler map, and the functional is computed from its detections or, for unresolved scatterers, its integrated cells. Sense-then-compute resolves detail that the functional discards, costing samples, data, and computation, and every integrated cell adds its own noise. Detection merges unresolved members, and millimeter-wave people counters degrade above two people per square meter \cite{toha2025mmcounter}.

Computing without resolving the contributors has a precedent in communications: in over-the-air computation (AirComp), a receiver obtains a function of concurrently transmitted data from their superposition without decoding them \cite{nazer2007computation,zhu2020broadband}, with federated learning as a leading application \cite{azimi2025airfl}. The function can also be extracted by structured lattice coding \cite{azimi2026aircpu} or read from the occupancy of shared resources without channel knowledge \cite{azimi2026occupancy}. In each case cooperating devices encode, precode, or schedule the signals that superpose. Integrated sensing and communication (ISAC) \cite{liu2022isac} and integrated sensing, communication, and computation (ISCC) \cite{wen2024survey}, even over the air \cite{li2023iscc}, compute on data that sensors have already measured. A scene offers neither cooperation nor measured data: its members are passive and unknown in number, position, and motion, and the only designable signal is the illumination.

Yet a scene can compute. For scatterers with uncorrelated returns, a correlator's expected readout energy sums the members' reflectivities, each weighted by the squared magnitude of the correlation, or cross-ambiguity function, of the receive template with the transmit waveform shifted by that member's delay and Doppler. This is the scattering-function relation of distributed-target theory \cite{vantrees1971detection}, used in clutter suppression \cite{stutt1968best,kay2007optimal}, scattering-function estimation \cite{gaarder1968scattering}, and weather radar \cite{doviak1993doppler}, but not to compute chosen functionals. The physics performs the sum, up to speckle (signal fluctuation) and noise; waveform and template choose the weights (the kernel), whose delay--Doppler integral is fixed by the transmit energy. We call this over-the-scene computation: the illumination is the program, the scene the computer, and the receiver reads only the result.

\begin{figure}[t]\centering\includegraphics[width=\columnwidth]{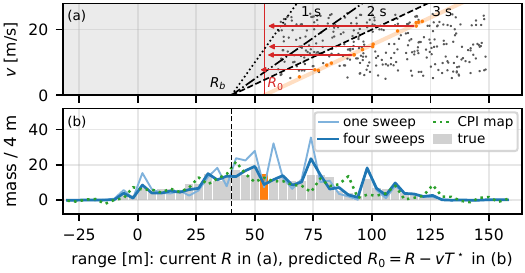}
\caption{Predicted range profiles (Theorem~\ref{thm:profile}) on the scene of Section~\ref{sec:numerical}. (a) Members by current range $R$ and speed $v$: after $3$~s the orange ones are at about $54$~m (arrows), so a sweep of horizon $\Tst=3$~s sums them in one $4$~m band of $R-v\Tst$; members left of the line of slope $1/\Tst$ through $(R_b,0)$ ($\Tst=1,2,3$~s) reach the boundary $R_b$ within $\Tst$ (shaded: $3$~s). (b) Profiles at $\Tst=3$~s and $10$~dB member signal-to-noise ratio (SNR) (orange: the band of (a)); dashed: $R_b$.}\label{fig:profiles}\end{figure}
A linear sweep computes how much is predicted to cross a boundary. At rate $\alpha$, dechirping (mixing the echo with the transmitted sweep) turns a member at range $R$ closing at speed $v$ into a beat tone of frequency $2\alpha R/\cc-2v/\lambda=2\alpha(R-v\Tst)/\cc$, with the \emph{horizon} $\Tst=\cc/(\lambda\alpha)$ ($\cc$ the speed of light, $\lambda$ the wavelength). This range--Doppler coupling makes the Doppler shift mimic a range offset of $-v\Tst$, the member's own displacement over $\Tst$ seconds, so each bin of the beat signal's discrete Fourier transform (DFT) collects the members predicted to share a range $\Tst$ seconds ahead. At $\Tst=3$~s, members at $100$~m closing at $20$~m/s and at $70$~m closing at $10$~m/s share a bin, both reaching $40$~m (the $3$~s line in Fig.~\ref{fig:profiles}(a)). The sum of the noise-floor-subtracted bins whose predicted range lies inside a boundary therefore estimates the reflectivity-weighted count, or mass, of the members predicted to have crossed it, with one noise term per bin rather than per map cell. Because $\Tst\propto1/\alpha$, a $3$~s horizon needs a sweep about $150$ times slower than a range--Doppler map's $36$~\textmu s chirps, whose horizon is only $20$~ms.

Aggregating members in one bin has a price: their echoes add with random phases, so the bin's energy has the relative fluctuation of a single echo, whereas resolving members averages their fluctuations at the cost of one noise term per cell. Aggregation therefore suits weak, noise-limited members and resolution strong, speckle-limited ones. Programming the illumination raises three questions: which functionals of an unknown, unresolvable population the scene can compute, with which waveform and at what noise cost, and whether the result can beat every unbiased estimator that resolves the population first. We answer them for general narrowband waveforms and then for the chirp.

\subsection{Contributions}
\begin{enumerate}
\item \emph{Framework and computability} (Section~\ref{sec:model}, Theorem~\ref{thm:transfer}). We formalize over-the-scene computation by programs, weighted sums of floor-subtracted readout energies whose means are kernel-weighted sums of member reflectivities. Through a classical identity for Cohen's-class time--frequency distributions \cite{cohen1989time,hlawatsch1992linear,werner1984quantum}, we show that the transmit ambiguity function is the transfer function of this computation: it determines which kernels a waveform can compute at finite noise cost and the minimal cost, bounded below over all waveforms.
\item \emph{Aggregation geometry} (Theorems~\ref{thm:gram}--\ref{thm:line}). One readout gathers a population's whole mass only if the members' echo signals are collinear. With Lieb's concentration bounds \cite{lieb1990integral} and the Brunn--Minkowski inequality \cite{gardner2002brunn}, we show that any delay--Doppler region pairwise aligned at correlation loss $\varepsilon$ has area at most about $0.68\varepsilon$ for small $\varepsilon$, whatever the waveform's duration and bandwidth, whereas along a line the aligned length can grow with duration and is maximal, among constant-envelope waveforms, for the chirp of the line's slope.
\item \emph{Chirp realization} (Theorem~\ref{thm:profile}, Propositions~\ref{prop:cumulative}--\ref{prop:moments}). We show that, by range--Doppler coupling \cite{fitzgerald1974effects}, a sweep's expected beat spectrum is the population's predicted range profile $\Tst$ seconds ahead, without Doppler processing or tracking; cumulative mass, flux, clutter-free profiles, and the mean and spread of closing speeds follow.
\item \emph{Statistics and limits} (Theorems~\ref{thm:variance} and \ref{thm:limits}, Proposition~\ref{prop:granularity}). Exact variances show when aggregation beats map integration (summing map cells), and, for unbiased estimation, a Cram\'er--Rao bound over all equal-energy waveforms places the switch from aggregation to resolution at a member signal-to-noise ratio (SNR) of one, with weaker members best grouped into unit-SNR sets. In the low-SNR limit, a sum of bin readouts beats every unbiased estimator from an equal-energy CPI whenever it has fewer terms than the CPI has independently fluctuating effective modes (Corollary~\ref{cor:every}).
\item \emph{Extremal functionals} (Theorems~\ref{thm:impossible} and \ref{thm:bisection}). Under Rayleigh fluctuation, no finite set of waveform--template probes, adaptive or not, determines an unconstrained population's first arrival, even without noise; adaptive bisection localizes it with logarithmically many tests when the first arrivals outweigh later arrivals' sidelobes.
\item \emph{Design and validation} (Sections~\ref{sec:moments}, \ref{sec:resources}, and \ref{sec:numerical}). We give the complete procedure, which needs up to 77 times fewer samples per horizon than a CPI of equal dwell and bandwidth, is 40 times less biased by noise-floor errors than map integration, and converts mass into counts for flocks of similar speeds. We validate the theory by simulation at 77~GHz under fluctuation, calibration error, path loss, and clustering.
\end{enumerate}

\section{Related Work}\label{sec:related}
\textbf{Task-oriented sensing and analog computing.} Task-specific-information waveform design \cite{kim2010waveform}, semantic sensing with an information-bottleneck objective \cite{zhang2026semantic}, and receive beamforming that minimizes the position error normal to a semantic boundary \cite{mojahedian2026task} shape the measurement for a decision on one target or user through its parameters. In a dense population, the number of parameters is unknown, and the Fisher information for separating neighbors vanishes as they merge. Learned metasurface illumination \cite{delhougne2020learned} and task-based quantization \cite{shlezinger2019hardware} optimize acquisition for a given task without characterizing which functionals it can compute. Wave-based analog computing lets the physics compute: a designed metamaterial applies an operator to data carried by an incident wave \cite{silva2014performing}; here the scene is data and computer, the wave the program. Imaging comes closer: semiparametric incoherent imaging estimates moments of unresolved sources by designing only the receiver \cite{tsang2019semiparametric}, and feature-specific structured \cite{baheti2006feature} and single-pixel imaging \cite{duarte2008single} measure inner products of a static scene with almost arbitrary patterns. Illumination of moving scatterers, by contrast, gives each probe only a squared cross-ambiguity surface; Theorem~\ref{thm:transfer} characterizes the kernels their combinations compute.

\textbf{Distributed targets and population sensing.} Clutter suppression shapes the receive filter \cite{stutt1968best} or transmit spectrum \cite{kay2007optimal} against the clutter predicted by the scattering-function relation, and scattering-function estimation inverts it: a matched-filter bank sees this function smoothed by the squared ambiguity function \cite{gaarder1968scattering}, and operator sampling \cite{pfander2006measurement} extends the estimation to overspread operators \cite{pfander2015estimation}. Here the scattering function is the scene measure (Section~\ref{sec:model}), which programs need not identify; Theorem~\ref{thm:transfer} gives the noise cost of the functionals they compute. Weather radar and radar aeroecology turn the reflectivity of unresolved scatterers into spectral moments and migration traffic rates \cite{doviak1993doppler,chilson2012partly}, crowd counting uses detections or features of standard waveforms \cite{toha2025mmcounter}, and the probability hypothesis density filter propagates the multitarget intensity, akin to the scene measure, from unassociated detections \cite{mahler2003multitarget}; all compute from cells or detections of a fixed illumination.

\textbf{Range--Doppler coupling.} Fitzgerald showed that a chirp measures the range of one tracked target at a shifted time \cite{fitzgerald1974effects}. Chirps of several rates serve as tomographic projections for imaging a range--Doppler reflectivity \cite{bernfeld1984chirp,feig1986tomographic}, and up--down sweeps separate a target's range and speed by pairing its beat peaks \cite{stove1992linear}. Here one projection, read without imaging, is a population's predicted range profile up to bounded errors (Theorem~\ref{thm:profile}), and several give kinematic moments without pairing (Proposition~\ref{prop:moments}).

\section{System Model and the Computation Identity}\label{sec:model}
Signals are in $L^2(\R)$ with $\ip{x}{y}=\int x(t)y^*(t)\,dt$ and $\norm{x}^2=\ip{x}{x}$. A scatterer at delay $\tau$ and Doppler shift $\nu$ acts, in the narrowband model, through the time--frequency shift $(\Tsh_\xi x)(t)=x(t-\tau)e^{j2\pi\nu t}$, $\xi=(\tau,\nu)\in\R^2$. The \emph{scene measure} is ${\mu=\sum_{k=1}^{K}\rho_k\delta_{\xi_k}}$, $\rho_k>0$, with $K$ unknown and $\rho_k$ the mean received energy of member $k$ per unit transmit energy (its reflectivity), proportional by the radar equation to its mean radar cross section times the two-way antenna gain and the path-loss factor ($\propto R_k^{-4}$ at range $R_k$). We call ${\mu(\sS)=\sum_{k\in\sS}\rho_k}$ the \emph{mass} of a set $\sS$ of members, which Section~\ref{sec:moments} converts into a plain count. A continuum of uncorrelated scatterers is a general finite measure, whose delay marginal is, up to gain and range scaling, what weather radar and radar aeroecology report as reflectivity \cite{doviak1993doppler,chilson2012partly}; angle enters as a beam factor (Section~\ref{sec:discussion}). A \emph{probe} is a pair $(s,h)$ of a transmit waveform of energy $E=\norm{s}^2$ and a unit-norm receive template. One look, a transmission of $s$, returns $r=\sum_{k=1}^{K}a_k\Tsh_{\xi_k}s+n$, with amplitudes $a_k$ and circular white Gaussian noise $n$ of density $\sigma^2$; the \emph{readout} is
\begin{equation}\label{eq:readout}
z=\ip{r}{h}=\sum_{k=1}^{K}a_k\,\chi_{s,h}(\xi_k)+\ip{n}{h},
\end{equation}
with the cross-ambiguity function $\chi_{s,h}(\xi):=\ip{\Tsh_\xi s}{h}$, the ambiguity function $\chi_s=\chi_{s,s}$, and $\ip{n}{h}\sim\mathcal{CN}(0,\sigma^2)$, the circular complex Gaussian law. The \emph{kernel} is $g_{s,h}=|\chi_{s,h}|^2$.
\begin{assumption}[Incoherence]\label{as:incoherence}
$\E a_k=0$, $\E a_ka_l^*=\rho_k\delta_{kl}$, and the amplitudes are independent of $n$.
\end{assumption}
\begin{assumption}[Block Rayleigh fluctuation]\label{as:gaussian}
$a_k\sim\mathcal{CN}(0,\rho_k)$, constant within a look and independent across members and looks and of the noise, with $\mu$ fixed across looks.
\end{assumption}
Assumption~\ref{as:incoherence} is the distributed-target model \cite{doviak1993doppler}: an echo's carrier phase $4\pi R_k/\lambda$ is uniform unless $R_k$ is known to a fraction of $\lambda/2$ ($2$~mm at 77~GHz). Assumption~\ref{as:gaussian}, used only for distributional statements, is the Swerling~I model with each look (sweep or CPI) as one scan \cite{swerling1960probability}. Within a look, the response $\Tsh_{\xi_k}s$, not $a_k$, carries the Doppler phase evolution. Independent renewal between looks, which lets them average speckle, models decorrelation by relative motion: members whose speeds differ by $\Delta v$ change their relative phase by $\pi$ within $\lambda/(4\Delta v)$, and vibration or sway decorrelates the rest. Rigidly coupled members keep Assumption~\ref{as:incoherence}, not the renewal (frozen amplitudes, Section~\ref{sec:numerical}).
\begin{proposition}[Computation identity]\label{prop:identity}
Under Assumption~\ref{as:incoherence}, $\E|z|^2=\int g_{s,h}\,d\mu+\sigma^2=\sum_k\rho_k|\chi_{s,h}(\xi_k)|^2+\sigma^2$.
\end{proposition}
Proposition~\ref{prop:identity}, the scattering-function relation for a discrete scene, holds because the cross terms vanish. A probe thus selects the linear functional $\mu\mapsto\int g_{s,h}\,d\mu$, which the receiver reads as the mean readout energy above the known floor $\sigma^2$ and estimates as in Lemma~\ref{lem:readout}.
\begin{lemma}[Readout law]\label{lem:readout}
Under Assumption~\ref{as:gaussian}, $|z|^2$ is exponential with mean $m+\sigma^2$, $m=\int g_{s,h}\,d\mu$, so with $M$ looks $\hat F=M^{-1}\sum_i|z_i|^2-\sigma^2$ is unbiased for $m$ with $\Var\hat F=(m+\sigma^2)^2/M$, the readouts' Cram\'er--Rao bound.
\end{lemma}

A \emph{program} $\sP=\{(s_i,h_i,q_i)\}_i$ is a finite family of probes with real weights. Its kernel is $g_\sP=\sum_iq_ig_{s_i,h_i}$, and the \emph{programmed readout} $\hat F_\sP=\sum_iq_i\hat F_i$, with $\hat F_i$ from Lemma~\ref{lem:readout}, is unbiased for $\int g_\sP\,d\mu$ whatever the number and positions of the members. The program \emph{computes} the functional $\int\Psi\,d\mu$ if $\Psi=g_\sP$: exactly in expectation, while finitely many looks estimate it with speckle and noise (Section~\ref{sec:statistics}); Theorem~\ref{thm:transfer} extends this to limits of one-waveform programs. The motivating functional is of this type up to smoothing: with $\tau=2R/\cc$ and $\nu=2v/\lambda$ for range $R$ and closing speed $v$, the mass predicted to lie inside a boundary $R_b$ after $\Tst$ seconds is $\int\Psi\,d\mu$ with $\Psi=\mathbf 1\{R-v\Tst\le R_b\}$, constant along delay--Doppler lines of slope $\alpha=\cc/(\lambda\Tst)$, the ridges of a chirp of rate $\alpha$. Since $\int\Psi\,d\mu$ ignores $\Psi$ where no member can lie, $\Psi$ may be tapered there, slowly along the lines, into $L^2(\R^2)$; its jump across the lines is computable only smoothed (Section~\ref{sec:align}).

\section{What Illumination Can Make the Scene Compute}\label{sec:compute}
The transmitter fixes one waveform per look, while the receiver may combine the readouts of any templates; the question is which kernels a waveform admits and at what receiver-noise cost.
\subsection{Conservation and the Transfer Theorem}\label{sec:transfer}
\begin{lemma}[Moyal \cite{moyal1949quantum,grochenig2001foundations}; Lieb \cite{lieb1990integral}]\label{lem:moyal}
For all $s,h,s_1,h_1,$ $s_2,h_2\in L^2(\R)$ and $p\ge2$,
\begin{align}
&\textstyle\iint\chi_{s_1,h_1}\chi_{s_2,h_2}^{*}\,d\tau\,d\nu=\ip{s_1}{s_2}\ip{h_1}{h_2}^{*},\label{eq:moyal}\\
&\textstyle\iint|\chi_{s,h}|^p\,d\tau\,d\nu\le\frac{2}{p}\norm{s}^p\norm{h}^p,\label{eq:lieb}
\end{align}
with equality in \eqref{eq:lieb} for matched Gaussians.
\end{lemma}
For $s_1=s_2$, $h_1=h_2$, \eqref{eq:moyal} gives the volume invariance of the ambiguity function \cite{vantrees1971detection}: \emph{every probe kernel has volume $\iint g_{s,h}=E$, the transmit energy}, and \eqref{eq:lieb} limits its concentration. A kernel of volume $V$ thus costs transmit energy $V$ or, on one transmission of energy $E$, $V/E$ orthonormal unit-weight templates, each adding a noise term (Theorem~\ref{thm:variance}).

Three representations settle which kernels are computable: writing the receiver as one operator $Q$ makes its kernel linear in $Q$; expanding $Q$ in time--frequency shifts gives its spreading function, whose energy is the noise cost; and in the symplectic Fourier domain the kernel becomes the product of this spreading function and the ambiguity function. We write programs as quadratic receivers: Hermitian forms of finitely many readouts $z_i=\ip{r}{h_i}$, $\ip{Qr}{r}=\sum_{i,j}q_{ij}z_jz_i^{*}$ with $Q=\sum_{i,j}q_{ij}h_ih_j^{*}$, $q$ Hermitian, and $h_ih_j^{*}f:=h_i\ip{f}{h_j}$; a one-waveform program is $Q=\sum_iq_ih_ih_i^*$, and every Hermitian $Q$ is one in its eigenbasis. By Assumption~\ref{as:incoherence}, $\E\ip{Qr}{r}=\int g_{s,Q}\,d\mu+\sigma^2\tr Q$ with $g_{s,Q}(\xi)=\ip{Q\Tsh_\xi s}{\Tsh_\xi s}$, so the kernel depends on the receiver only through $Q$. The receiver's counterpart of $\chi_s$ is its \emph{spreading function} $\eta_Q=\sum_{i,j}q_{ij}\chi_{h_j,h_i}^*$, which expands $Q$ in time--frequency shifts, $\ip{Qf}{f'}=\iint\eta_Q(\zeta)\ip{\Tsh_\zeta f}{f'}\,d\zeta$ with $\zeta=(x,\omega)$. Its energy $\norm{\eta_Q}_{L^2}^2=\norm{Q}_{\HS}^2$, the squared Hilbert--Schmidt norm \cite{grochenig2001foundations}, is $\sum_iq_i^2$ for a program with orthonormal templates. The \emph{symplectic Fourier transform} $(\Fs F)(\tau,\nu)=\iint F(x,\omega)e^{j2\pi(\omega\tau-\nu x)}dx\,d\omega$ is a unitary involution.
\begin{theorem}[Computation transfer]\label{thm:transfer}
Let $s\in L^2(\R)$ have energy $E$ and let $Q$ be a finite-rank Hermitian receiver.
\begin{enumerate}
\item[(i)] $\Fs g_{s,Q}=\eta_Q\,\chi_s$; equivalently, $g_{s,Q}=\Fs[\eta_Q\chi_s]$.
\item[(ii)] Define the \emph{noise cost} of a real kernel $\Psi\in L^2(\R^2)$ as
\begin{equation}\label{eq:cost}
c_s(\Psi)=\iint_{\{\chi_s\neq0\}}\frac{|\Fs\Psi|^2}{|\chi_s|^2}\,d\zeta
\end{equation}
if $\Fs\Psi=0$ almost everywhere (a.e.) on $\{\chi_s=0\}$, and $c_s(\Psi)=\infty$ otherwise. Then $\Psi$ is \emph{computable} with $s$ (a uniform and $L^2$ limit of kernels $g_{s,Q_n}$ of finite-rank Hermitian receivers $Q_n$ with bounded Hilbert--Schmidt norms) if and only if $c_s(\Psi)<\infty$, and the $Q_n$ can be chosen with $\norm{Q_n}_{\HS}^2\le c_s(\Psi)$; every such sequence has $\liminf_n\norm{Q_n}_{\HS}^2\ge c_s(\Psi)$, and every receiver with $g_{s,Q}=\Psi$ has $\norm{Q}_{\HS}^2\ge c_s(\Psi)$.
\item[(iii)] Under Assumption~\ref{as:gaussian}, $\hat F_Q=M^{-1}\sum_{i=1}^M(\ip{Qr_i}{r_i}-\sigma^2\tr Q)$, with $r_i$ the return of look $i$, is unbiased for $\int g_{s,Q}\,d\mu$ and $\Var\hat F_Q\ge\sigma^4\norm{Q}_{\HS}^2/M$; hence every receiver with $g_{s,Q}=\Psi$ has $\Var\hat F_Q\ge\sigma^4c_s(\Psi)/M$.
\item[(iv)] $c_s(\Psi)\ge\norm{\Fs\Psi}_{L^1}^2/E^2$ for every waveform of energy $E$, with equality, if finite, exactly when $|\Fs\Psi|\propto|\chi_s|^2$ a.e.
\end{enumerate}
\end{theorem}
Part (i) gives a classical identity a new role: $g_{s,Q}(\xi)=\ip{\Tsh_\xi^*Q\Tsh_\xi s}{s}$ is, up to a reflection of the plane, the Cohen's-class distribution of $s$ generated by $Q$, and (i) is its ambiguity-domain form \cite{cohen1989time,hlawatsch1992linear}, Werner's convolution theorem \cite{werner1984quantum}. In time--frequency analysis this distribution is displayed; here an unknown scene integrates it. The transmitter fixes $\chi_s$, the receiver chooses $\eta_Q$, and the physics multiplies them in the symplectic domain: \emph{the transmit ambiguity function is the transfer function of the scene computer}. By (ii), a waveform computes, as bounded-noise limits of programs, exactly the kernels whose inverse filter $\eta_Q=\Fs\Psi/\chi_s$ has finite energy; the eigenvectors and eigenvalues of the receiver with this spreading function are the programs' templates and weights, and by (iii) its energy $c_s(\Psi)$ sets the receiver-noise term of the variance, dominant at low SNR. Finer detail requires larger lags: a change over a Doppler (delay) interval $1/\Delta$ puts $\Fs\Psi$ at delay (Doppler) lags of order $\Delta$, so a waveform computes only detail at lags its ambiguity function reaches, and design reduces to placing $|\chi_s|$ where $|\Fs\Psi|$ is large (see \eqref{eq:chirpcost} for the chirp); part (iv) bounds what any waveform can achieve, with equality for the matched kernel $\Psi\propto|\chi_s|^2$, extending matched illumination \cite{bell1993information} to computation.

\subsection{Aggregation in One Readout}\label{sec:gram}
Theorem~\ref{thm:transfer} gives the noise cost of a kernel; the signal is the mass that one template captures.
\begin{theorem}[Aggregation in one readout]\label{thm:gram}
Let $u_k=\Tsh_{\xi_k}s$ be the members' responses and $\Gamma_s=\sum_k\rho_ku_ku_k^{*}$ the \emph{echo covariance operator}. For every unit-norm $h$,
\begin{equation}\label{eq:gram}
\int g_{s,h}\,d\mu=\ip{\Gamma_sh}{h}\le\vartheta_{\max}(\Gamma_s)\le E\,\mu(\R^2),
\end{equation}
with $\vartheta_{\max}$ the largest eigenvalue, equality on the left for its eigenvector, and on the right if and only if all members' responses are collinear. (i) If $|\ip{u_k}{u_l}|^2\ge(1-\varepsilon)E^2$ for all $k,l$ in a subset $\sS$ (pairwise alignment at loss $\varepsilon$), the template $u_{k_0}/\sqrt E$ of any $k_0\in\sS$ captures at least $(1-\varepsilon)E\mu(\sS)$; (ii) if the responses in $\sS$ are orthogonal, every unit-norm $h$ captures at most $E\max_{k\in\sS}\rho_k$ of their mass.
\end{theorem}
An illumination that aligns a subpopulation's responses lets one readout capture nearly its whole mass with one noise term, even though the uncorrelated amplitudes add only in mean energy; one that resolves the members caps a readout at one member's worth. The waveform, not the receiver, thus caps the readout SNR of an aggregate through the \emph{geometry of the response set}; alignment costs speckle diversity (Section~\ref{sec:statistics}).

\subsection{Alignment Geometry and Optimal Lines}\label{sec:align}
Since $|\ip{\Tsh_\xi s}{\Tsh_{\xi'}s}|=|\chi_s(\xi-\xi')|$ (Lemma~\ref{lem:weyl}), the ambiguity function decides which members one waveform can aggregate.
\begin{theorem}[Alignment geometry]\label{thm:geometry}
Let $s$ have energy $E$, $\varepsilon\in(0,1)$, $\mathcal Z_\varepsilon(s)=\{\zeta:|\chi_s(\zeta)|^2\ge(1-\varepsilon)E^2\}$, and
\begin{equation}\label{eq:aeps}
a(\varepsilon)=\min_{p\ge2}\frac{2}{p}(1-\varepsilon)^{-p/2}=
\begin{cases} e\,\ln\frac{1}{1-\varepsilon}, & \varepsilon\le 1-e^{-1},\\ (1-\varepsilon)^{-1}, & \text{otherwise}.\end{cases}
\end{equation}
(i) The area $|\mathcal Z_\varepsilon(s)|\le a(\varepsilon)$. (ii) If a compact $\mathcal A\subset\R^2$ is pairwise aligned at loss $\varepsilon$, i.e., $|\chi_s(\xi-\xi')|^2\ge(1-\varepsilon)E^2$ for all $\xi,\xi'\in\mathcal A$, then $\mathcal A-\mathcal A\subset\mathcal Z_\varepsilon(s)$ and $|\mathcal A|\le a(\varepsilon)/4$. (iii) For the chirp $s(t)=e^{j\pi\alpha t^2}$ on $[-T/2,T/2]$ and $|\tau|\le T$,
\begin{equation}\label{eq:chirpaf}
\frac{|\chi_s(\tau,\nu)|^2}{E^2}=\Big(1-\frac{|\tau|}{T}\Big)^2\sinc^2\!\big((\nu-\alpha\tau)(T-|\tau|)\big),
\end{equation}
$\sinc x=\sin(\pi x)/(\pi x)$, a ridge along $\nu=\alpha\tau$ \cite{vantrees1971detection}.
\end{theorem}
Delay--Doppler area is dimensionless, so (ii) bounds the time--bandwidth product of any pairwise-aligned region by $a(\varepsilon)/4$, about $0.68\varepsilon$ for small $\varepsilon$, whatever the waveform's own duration and bandwidth. Aggregation also costs resolution:
\begin{corollary}[Computation--resolution trade-off]\label{cor:tradeoff}
If $\mathcal A\subset\R^2$ is pairwise aligned at loss $\varepsilon$, then $|g_{s,h}(\xi)-g_{s,h}(\xi')|\le E\sqrt\varepsilon$ for every unit-norm $h$ and $\xi,\xi'\in\mathcal A$, so scenes $\mu_1,\mu_2$ on $\mathcal A$ of equal mass give mean readout energies within $E\mu_1(\mathcal A)\sqrt\varepsilon$.
\end{corollary}
Lines have zero area, but members predicted to share a range lie on one (Section~\ref{sec:model}); Theorem~\ref{thm:line} bounds a line's aligned length. Write $s_\alpha(t)=s(t)e^{-j\pi\alpha t^2}$, the waveform dechirped at rate $\alpha$.
\begin{theorem}[Optimal alignment on lines]\label{thm:line}
Let $s$ have energy $E$ and support in an interval of length $T$.
(i) $\chi_s(\Delta,\alpha\Delta)=e^{j\pi\alpha\Delta^2}\int s_\alpha(t-\Delta)s_\alpha^*(t)\,dt$, so a segment of a line of slope $\alpha$ with delay length $L$ is pairwise aligned at loss $\varepsilon$ if and only if $|\chi_s(\Delta,\alpha\Delta)|\ge\sqrt{1-\varepsilon}\,E$ for $|\Delta|\le L$.
(ii) If $|s|=\sqrt{E/T}$ on an interval of length $T$ and $s=0$ elsewhere (constant envelope), then $|\chi_s(\Delta,\alpha\Delta)|\le E\max\{1-|\Delta|/T,0\}$, so $L\le T(1-\sqrt{1-\varepsilon})$. The linear chirp $\sqrt{E/T}\,e^{j(\pi\alpha t^2+2\pi f_0t+\phi_0)}$, with any $f_0$ and $\phi_0$, attains equality for every $|\Delta|\le T$, hence the maximal $L$ at every loss, and is the only continuous-phase constant-envelope waveform attaining it for all $\Delta$ in an interval $(0,\Delta_0)$.
\end{theorem}
A constant-envelope illumination of duration $T$ thus aligns at most a segment of length about $\varepsilon T/2$ on a line. The linear chirp attains this on its own lines, and alignment along them is nearly lossless, since a population's delay spread (microseconds) is far below the millisecond frequency-modulated continuous-wave (FMCW) sweeps of Section~\ref{sec:numerical}.

Applied to the chirp, Theorem~\ref{thm:transfer}(ii) delimits the kernels it computes cheaply. By \eqref{eq:chirpaf}, $|\chi_s(x,\alpha x+\omega')|^2\le E^2/(\pi\omega'T)^2$ at an offset $\omega'\ne0$ from the ridge, and multiplication by $\omega-\alpha x$ in the symplectic domain is the derivative $(\partial_\tau+\alpha\partial_\nu)/(j2\pi)$ along the lines; Parseval's identity and \eqref{eq:cost} then give
\begin{equation}\label{eq:chirpcost}
c_s(\Psi)\ge\Big(\frac{T}{2E}\Big)^2\norm{(\partial_\tau+\alpha\partial_\nu)\Psi}^2_{L^2}.
\end{equation}
Rapid kernel variation along the chirp's lines, parallel to its ambiguity ridge, is thus costly: a kernel of a bin's height and width changing by a factor of order one within a delay interval $\Delta\tau$ along a line $\nu-\alpha\tau=\text{const}$ costs noise of order $T/\Delta\tau$ bin readouts or more. Affordable kernels therefore vary along the lines only on delay scales of about $T$ or more, so over the population's far shorter delay spread they are functions of $\nu-\alpha\tau$. Across the lines, $\chi_s=0$ for $|x|\ge T$ limits every computable kernel to Doppler scales of $1/T$ or more, a predicted range of $\cc/(2B)$ for the bandwidth $B=|\alpha|T$, so a sharp boundary is computed only smoothed to this scale (Theorem~\ref{thm:profile}).

\section{The Chirp Computes the Predicted Range Profile}\label{sec:chirp}
By \eqref{eq:chirpcost}, a chirp's affordable kernels are functions of predicted range (Section~\ref{sec:model}): one sweep estimates the masses of predicted-range sets; several yield flux, moments, and counts.
\subsection{Dechirp Model and the Predicted Profile}
An FMCW transmitter radiates on $[0,T)$ the chirp $s(t)=\exp\{j2\pi(f_ct+\alpha t^2/2)\}$, with energy $E=T$, carrier frequency $f_c=\cc/\lambda$, chirp rate $\alpha$, and bandwidth $B$. Member $k$ has range $R_k-v_kt$ with closing speed $v_k$, round-trip delay $\tau_k(t)=\tau_k-\beta_kt$, $\tau_k=2R_k/\cc$, $\beta_k=2v_k/\cc$. Dechirping, $\tilde r=s\,r^*$ after low-pass filtering, gives the beat signal (Appendix~\ref{app:chirp})
\begin{align}\label{eq:beat}
\tilde r(t)&=\sum_k a_k^*e^{j\theta_k}\,e^{j2\pi f_kt}\,e^{-j2\pi\kappa_kt^2}+\tilde n(t),\nonumber\\
f_k&=\alpha\tau_k-\nu_k,\quad \nu_k=\frac{2v_k}{\lambda},\quad \kappa_k=\frac{2\alpha v_k}{\cc},
\end{align}
up to relative corrections of order $\beta_k\sim10^{-7}$. Here $\kappa_k$, from motion during the sweep, drifts the beat tone by $2\kappa_kT$, at most $4/\gamma$ bins for the margin $\gamma$ of Assumption~\ref{as:stophop}; $\theta_k$ are constant phases, and $\tilde n$ is white noise of density $\sigma^2$; the $a_k^*$ obey Assumptions~\ref{as:incoherence} and \ref{as:gaussian}. The bin readout is $z_b=\int_0^T\tilde r(t)w(t)e^{-j2\pi f_bt}dt$ with a real unit-norm window $w$ on $[0,T]$, so $z_b^*=\ip{r}{h_b}$ is the readout \eqref{eq:readout} of the template $h_b=s\,w\,e^{-j2\pi f_bt}$. For the rectangular window $w=T^{-1/2}\mathbf 1_{[0,T]}$ and $f_b=b/T$, these are the DFT bins, with orthonormal templates.
\begin{assumption}[Narrowband sweep]\label{as:stophop}
$B\ll f_c$, $\tau_k\ll T$, constant speeds over the sweep, and the stop-and-hop condition $\max_k|v_k|T\le\cc/(\gamma B)$ for a margin $\gamma>0$: a member moves at most $2/\gamma$ of a range bin $\cc/(2B)$ per sweep.
\end{assumption}
\begin{theorem}[Predicted range profile]\label{thm:profile}
For the beat model \eqref{eq:beat} with echo onsets $\tau_k$, under Assumptions~\ref{as:incoherence} and \ref{as:stophop},
\begin{align}\label{eq:profile}
\E|z_b|^2&=\sum_k\rho_k|W_k(f_b)|^2+\sigma^2,\nonumber\\
W_k(f)&=\int_{\tau_k}^Tw(t)e^{j2\pi(f_k-f)t-j2\pi\kappa_kt^2}dt,
\end{align}
with $|W_k(f_b)-\widehat W(f_b-f_k)|\le\norm{w}_\infty(\frac{4\pi}{3\gamma}T+\tau_k)$, ${\widehat W(f)=\int w(t)e^{-j2\pi ft}dt}$, the term $\tau_k$ accounting for the echo onset. The ideal kernel $|\widehat W(f_b-f_k)|^2$ is a ridge of beat-frequency width of order $1/T$ along
\begin{align}
f_k=f_b&\iff R_k-v_k\Tst=R_0,\nonumber\\
\Tst&=\frac{\cc}{\lambda\alpha},\qquad R_0=\frac{\cc f_b}{2\alpha}:\label{eq:locus}
\end{align}
the mean energy of bin $f_b$ above the floor, divided by $T$, is the mass of all members whose linearly predicted range after $\Tst$ seconds is $R_0$, smoothed by the unit-sum kernel $|\widehat W|^2/T$ to a range resolution of order $\cc/(2B)$; up to the error above, the expected beat spectrum is the population's predicted range profile $\Tst$ seconds ahead.
\end{theorem}
A down-chirp gives $\Tst<0$, the profile $|\Tst|$ seconds in the past. Members of equal predicted range are aligned (Theorem~\ref{thm:gram}(i)) and unresolved (Corollary~\ref{cor:tradeoff}): the scene computes the linear extrapolation of its own motion. A member whose closing speed grows at rate $\dot v_k$ ends up $\dot v_k(\Tst)^2/2$ closer than predicted. Each member enters with its current reflectivity, not the one at its predicted range; Section~\ref{sec:moments} corrects the path loss for counts.

\subsection{Derived Functionals}
\begin{proposition}[Cumulative mass and flux]\label{prop:cumulative}
For a boundary range $R_b$, a lower limit $R_{\min}$, an up-chirp, and the DFT bins $\mathcal B$ with $R_{\min}\le R_0\le R_b$, let $C=\sum_{b\in\mathcal B}{(|z_b|^2-\sigma^2)}$. With $\kappa_k=0$ and the onset neglected, \eqref{eq:profile} gives $\E C=T\sum_k\rho_k\varpi_k$ with $\varpi_k=\sum_{b\in\mathcal B}\sinc^2{((f_b-f_k)T)}\in[0,1]$; for $d>1$, $1-\varpi_k\le2/(\pi^2(d-1))$ for a member whose predicted range is $d$ bins inside $[R_{\min},R_b]$, and $\varpi_k\le1/(\pi^2(d-1))$ for one $d$ bins outside.
\end{proposition}
The proof uses $\sum_b\sinc^2(b-x)=1$ and an integral bound on the tails. Thus, for $R_{\min}$ below all predicted ranges, $C/T$ estimates the motivating functional of Section~\ref{sec:model}, the mass predicted inside $R_b$ after $\Tst$, up to a bias from members near the boundary, and, for approaching members, its difference between two horizons estimates the mass predicted to cross $R_b$ between them, the integrated flux. Static scatterers ($\nu=0$) bias these masses at every horizon; a moving-target-indication (MTI) template removes them:
\begin{proposition}[MTI template]\label{prop:mti}
For two back-to-back sweeps, $T_p=T$ apart, with amplitudes constant over the pair, every bin of $(\tilde r_1-\tilde r_2)/\sqrt2$ obeys \eqref{eq:profile}, up to an offset of at most $2/\gamma$ of a bin between the two sweeps' kernels, with the kernel multiplied by $2\sin^2(\pi\nu_kT_p)$ and the same floor $\sigma^2$: static scatterers vanish, and a mover is weighted by a factor in $[0,2]$ whose Doppler average is one.
\end{proposition}

\subsection{Kinematic Moments and Counts}\label{sec:moments}
Let $p_{\Tst}(b)=(\E|z_b|^2-\sigma^2)/T$ be the expected profile of Theorem~\ref{thm:profile}, with centroid $\bar R(\Tst)$ and variance $V(\Tst)$ over $R_0$. It samples, after kernel smoothing, the distribution of $R-v\Tst$, a Radon projection of the joint range--speed distribution \cite{feig1986tomographic}.
\begin{proposition}[Kinematic moments]\label{prop:moments}
Under the conditions of Theorem~\ref{thm:profile} with $\kappa_k=0$, the echo onset neglected, the Hann window $w(t)=\sqrt{8/(3T)}\sin^2(\pi t/T)$, and all bins $f_b=b/T$, $b\in\mathbb Z$, every horizon has $\sum_bp_{\Tst}(b)=\mu(\R^2)$ and
\begin{align}\label{eq:moments}
\bar R(\Tst)&=\bar R-\Tst\,\bar v,\nonumber\\
V(\Tst)&=V_R-2\Tst V_{Rv}+(\Tst)^2V_v+\Delta_R^2/3,
\end{align}
where $\bar R$, $\bar v$, $V_R$, $V_v$, $V_{Rv}$ are the reflectivity-weighted means, variances, and covariance of ranges and closing speeds and $|\Delta_R|=\cc/(2B)$ is the bin spacing; $\Delta_R^2/3$ is the Hann kernel's own variance. Rectangular-window bins, summed symmetrically, give a position-dependent centroid bias of up to $|\Delta_R|/(2\pi)$ and generically infinite variance.
\end{proposition}
Hence the mean closing speed $\bar v=-d\bar R(\Tst)/d\Tst$ follows from an up--down pair of chirps, and a third horizon gives the second moments.

\textbf{From mass to counts.} Counting requires compensating each member's $R^{-4}$ path loss and dividing the mass by a mean cross section. The path loss depends on the current range $R_0+v\Tst$, which varies with $v$ within a bin, and weighting members individually would vary the kernel along a chirp's lines over the delay spread $\Delta\tau$ ($0.6$~\textmu s against $T=5.4$~ms) at a noise cost of order $T/\Delta\tau\approx10^4$ bin readouts by \eqref{eq:chirpcost}. Whole bins are therefore rescaled by $(R_0+\bar v\Tst)^4$, with $\bar v$ from Proposition~\ref{prop:moments}, at a relative error of about $4|\Tst|\,|v-\bar v|/R$ for a member at range $R$, which averages out when the speeds in each bin scatter around $\bar v$, as in a flock. Dividing by a calibrated mean cross section, as radar aeroecology does \cite{chilson2012partly}, leaves, for members with independent cross sections and beam gains, a relative error of about their coefficient of variation over the square root of the count.

\section{Statistics, Limits, and Comparison with Sense-Then-Compute}\label{sec:statistics}
\subsection{The Exact Variance of Readout Sums}
The masses of Section~\ref{sec:chirp} are sums of readout energies, and so is map integration: a CPI of short chirps of period $T_s$ followed by a two-dimensional DFT gives a map $|\ip{r}{h_{\xi_c}}|^2$ on cells $\xi_c$ with templates $h_{\xi_c}$, and integrating it along a line of $n$ cells of equal predicted range is a program with a kernel set by the map and floor $n\sigma^2$.
\begin{theorem}[Variance of readout sums]\label{thm:variance}
Let $h_1,\dots,h_n$ be orthonormal templates, $W\in\C^{n\times K}$ with $W_{bk}=\ip{u_k}{h_b}$, $D=\operatorname{diag}(\rho_k)$, and $G=D^{1/2}W^{\mathsf H}WD^{1/2}$ the \emph{captured Gram matrix}. Under Assumption~\ref{as:gaussian} with $M$ looks, $\hat F=M^{-1}\sum_{i,b}(|\ip{r_i}{h_b}|^2-\sigma^2)$ satisfies
\begin{equation}\label{eq:variance}
\E\hat F=\tr G,\quad \Var\hat F=\big(\norm{G}_F^2+2\sigma^2\tr G+n\sigma^4\big)/M,
\end{equation}
and, if $G\ne0$, $(\tr G)^2/\min(n,K)\le(\tr G)^2/\rank G\le\norm{G}_F^2\le(\tr G)^2$, with equality on the right iff $\rank G=1$.
\end{theorem}
The terms are speckle, signal--noise interaction, and noise. For a given captured mass $\tr G$, aggregation into one mode (rank one), an independently fluctuating echo component, maximizes the speckle, and splitting the mass into orthogonal modes diversifies it at the price of one $\sigma^4$ per template, the floor $\sigma^4\norm{Q}_{\HS}^2$ of Theorem~\ref{thm:transfer}(iii) for the projector $Q$ onto the templates. Independent circular non-Gaussian amplitudes add $\sum_k(\E|a_k|^4-2\rho_k^2)(\sum_b|W_{bk}|^2)^2$ to $M\Var\hat F$, a term negative for steady cross sections and only a fraction $(\E|a_k|^4/\rho_k^2-2)/K_g$ of the speckle term when each mode holds $K_g$ members of equal mass.
\begin{corollary}[Aggregate versus partition]\label{cor:regime}
Under Assumption~\ref{as:gaussian}, let the readout mass $m_\sS=E\mu(\sS)$ of a member set $\sS$ be read by one aligned template ($n=1$) or by $n\ge2$ orthonormal templates each capturing a disjoint part $m_b$ with full gain (exact for collinear groups, within a factor $1-\varepsilon$ at loss $\varepsilon$). Then $M\Var_{\mathrm{agg}}=(m_\sS+\sigma^2)^2$ and $M\Var_{\mathrm{part}}=\sum_bm_b^2+2m_\sS\sigma^2+n\sigma^4\ge m_\sS^2/n+2m_\sS\sigma^2+n\sigma^4$, so aggregation beats $n$ equal parts in variance if and only if $m_\sS/\sigma^2<\sqrt n$.
\end{corollary}
This is Theorem~\ref{thm:variance} for $G$ of rank one or block diagonal.

\subsection{Estimation Limits and Optimal Granularity}\label{sec:limits}
To see whether any processing or waveform can beat the estimators of Corollary~\ref{cor:regime}, we bound every unbiased estimator for a given waveform and over all waveforms. For a set $\sS$ of $|\sS|$ members let $\Gamma_{s,\sS}=\sum_{k\in\sS}\rho_ku_ku_k^*$, of trace $m_\sS$, have nonzero eigenvalues $\vartheta_i$, the masses of its orthogonal modes, and let $n_{\mathrm{eff}}(\sS)=(\tr\Gamma_{s,\sS})^2/\norm{\Gamma_{s,\sS}}_{\HS}^2\in[1,\rank\Gamma_{s,\sS}]$ be the effective number of modes into which $s$ splits the mass of $\sS$. Member $k$ has readout mass $m_k=E\rho_k$, and $m_k/\sigma^2$ is its SNR. The bounds hold even for known member positions.
\begin{theorem}[Estimation limits]\label{thm:limits}
Under Assumption~\ref{as:gaussian} with $M$ looks, every estimator $\hat m$ of $m_\sS$ that is a function of finitely many readouts of the returns and is unbiased for all reflectivities near $\rho=(\rho_k)$ satisfies, with $\psi(\vartheta)=\vartheta^2/(\vartheta+\sigma^2)^2$,
\begin{align}
\Var\hat m&\ge\frac{m_\sS^2}{M\sum_{i}\psi(\vartheta_i)}\ge\frac1M\max\Big\{\frac{m_\sS^2}{\rank\Gamma_{s,\sS}},\ n_{\mathrm{eff}}(\sS)\,\sigma^4\Big\},\label{eq:limit}\\
\Var\hat m&\ge\frac{m_\sS^2}{M\sum_{k\in\sS}\hat\psi(m_k)}\ge\frac1M\max\Big\{4m_\sS\sigma^2,\ \frac{m_\sS^2}{|\sS|}\Big\},\label{eq:universal}
\end{align}
where $\hat\psi(\vartheta)=\vartheta/(4\sigma^2)$ for $\vartheta\le\sigma^2$ and $\hat\psi(\vartheta)=\psi(\vartheta)$ otherwise. Bound \eqref{eq:universal} holds for every waveform of energy $E$, even one that changes from look to look.
\end{theorem}
For a given waveform, \eqref{eq:limit} extends the classical amplitude bound of a spread target \cite{vantrees1971detection} to a subset of the population. It has a speckle floor and a mode-count floor $n_{\mathrm{eff}}\sigma^4$, one noise term per effective mode, which dominates at low SNR. As the modes divide the mass ($\sum_i\vartheta_i=m_\sS$), the bound is lowest when each mode maximizes its information per unit mass, $\psi(\vartheta)/\vartheta=\vartheta/(\vartheta+\sigma^2)^2$ ($\psi(\vartheta)$ is a mode's information about its log-mass); this ratio peaks at unit SNR, the trade-off of Section~\ref{sec:intro} in exact form: a weaker mode spends its noise term on too little signal, and a stronger one carries speckle that splitting it would average out. Bound \eqref{eq:universal} holds for every waveform because a mode's information is at most a concave function $\hat\psi$ of its mass, and mode masses are never more even than member masses (the $\vartheta_i$ majorize the $m_k$). Whatever the illumination, the members contribute at most $\sum_k\hat\psi(m_k)$ to the information about the mass, and that much only if members above unit SNR are resolved and weaker ones aggregated into modes of unit SNR. Exact groups attain the bound:
\begin{proposition}[Efficient readouts and optimal granularity]\label{prop:granularity}
Under Assumption~\ref{as:gaussian}, the readout estimator $\hat F_Q$ of Theorem~\ref{thm:transfer}(iii) attains the Cram\'er--Rao bound of its target $\int g_{s,Q}\,d\mu$ if and only if $\Sigma Q\Sigma\in\operatorname{span}\{u_ku_k^*\}$ \cite{kay1993fundamentals}, where $\Sigma=\Gamma_s+\sigma^2I$ on the span of the responses and of the range of $Q$. In particular, if the responses of $\sS$ form groups that are collinear within and orthogonal across groups and to all other responses, the readout sum with one aligned template per group is unbiased for $m_\sS$ and efficient, with variance $\sum_\ell(m_\ell+\sigma^2)^2/M$, $m_\ell$ the mass of group $\ell$. Hence, for members of equal mass, \eqref{eq:universal} equals $(m_\sS+|\sS|\sigma^2)^2/(M|\sS|)$ if $m_k\ge\sigma^2$, attained when every member is resolved (groups of one), and $4m_\sS\sigma^2/M$ if $m_k\le\sigma^2$, attained, when $\sigma^2/m_k$ is an integer dividing $|\sS|$, by groups of aggregate mass $\sigma^2$.
\end{proposition}
Exact groups require co-location, but aligned members approach them within the loss $\varepsilon$ (Theorem~\ref{thm:gram}(i)). In an FMCW profile, members of equal predicted range do so with loss of order $\Delta\tau/T$, and members of one bin at beat offset $\Delta f$ keep correlation $|\sinc(\Delta f\,T)|$. For weak members in equal modes of SNR $x=m_\sS/(n_{\mathrm{eff}}\sigma^2)$, bound \eqref{eq:limit} equals the grouped variance and exceeds \eqref{eq:universal} by the factor ${(1+x)/(2\sqrt x)=\cosh(\ln\sqrt x)}$ in standard deviation. By this cosh law, the member SNR enters the gap only through $x$, and modes of unit SNR close it.
\begin{corollary}[Aggregation against every unbiased estimator]\label{cor:every}
Under Assumption~\ref{as:gaussian}, let $s$ and $s'$ have equal energy, let $n_{\mathrm{eff}}^{s'}(\sS)$ be the effective number of modes under $s'$, and let a programmed readout $\hat F$ under $s$ sum over $n$ orthonormal templates with captured Gram matrix $G$ and bias $\tr G-m_\sS$. If
\begin{equation}\label{eq:every}
n+\frac{2\tr G}{\sigma^2}+\frac{\norm{G}_F^2+M(\tr G-m_\sS)^2}{\sigma^4}<n_{\mathrm{eff}}^{s'}(\sS),
\end{equation}
the mean-squared error of $\hat F$ is below the variance of every unbiased estimator of $m_\sS$ from $M$ looks of $s'$, even one that knows the positions. As $\sigma^2\to\infty$, \eqref{eq:every} becomes $n<n_{\mathrm{eff}}^{s'}(\sS)$.
\end{corollary}
The corollary follows from Theorem~\ref{thm:variance}, $M\E(\hat F-m_\sS)^2=\norm{G}_F^2+2\sigma^2\tr G+n\sigma^4+M(\tr G-m_\sS)^2$, and from \eqref{eq:limit}, $M\Var\hat m\ge n_{\mathrm{eff}}^{s'}(\sS)\sigma^4$. For a resolving waveform, $n_{\mathrm{eff}}^{s'}(\sS)$ grows with the number of occupied cells, while $n$, the number of summed predicted-range bins, does not.

\subsection{Sense-Then-Compute and Resources}\label{sec:resources}
Sense-then-compute has two further costs. Detection followed by counting (detect-then-count) is neither a readout sum nor unbiased: a detector that reports at most one detection per cell counts at most $\min(n,K)$ of $K$ members in $n$ cells, an undercount of at least $1-n/K$ at every SNR.

The map itself costs samples. A chirp of rate $\alpha$ has the complex beat bandwidth $2\alpha R_{\max}/\cc+4v_{\max}/\lambda$ for a maximum range $R_{\max}$ and speed $v_{\max}$. A CPI must sample Doppler without ambiguity, $T_s\le\lambda/(4v_{\max})$, so its chirp rate $B/T_s$ is high; the programmed sweep forms no Doppler axis, and its rate $\alpha=\cc/(\lambda\Tst)$ is set by the horizon. Table~\ref{tab:resources} shows $77$ times fewer samples, with ratios of $66$--$77$ for horizons of $1$--$3$~s ($61$--$71$ at $T_s=\lambda/(4v_{\max})$) and still $35$--$47$ if all members close: one-sided Doppler halves the term $4v_{\max}/\lambda$ of both beat bandwidths and lets the CPI use $T_s=\lambda/(2v_{\max})$. The saving is per horizon, each needing its own sweep, at equal dwell, bandwidth, and energy. Sub-Nyquist radars undersample too but rely on sparsity \cite{barilan2014subnyquist}, which dense populations lack.
\begin{table}[t]\centering
\caption{Resources per look at the $\Tst=3$~\textnormal{s} design ($T=5.4$~\textnormal{ms}, $B=139$~\textnormal{MHz}, $R_{\max}=155$~\textnormal{m}, $v_{\max}=25$~\textnormal{m/s})}\label{tab:resources}
\setlength{\tabcolsep}{2.5pt}\footnotesize
\begin{tabular}{@{}lcc@{}}\toprule
 & programmed sweep & CPI ($T_s=36$~\textmu s)\\\midrule
chirps per look / chirp rate & 1 / 25.7~GHz/s & 150 / 3.85~THz/s\\
complex beat bandwidth & 52~kHz & 4.0~MHz\\
complex samples $N$ & 282 & 21\,667\\
transform work ($N\log_2N$) & $2.3\times10^3$ & $3.1\times10^5$\\
\bottomrule\end{tabular}
\end{table}

\section{Extremal Functionals}\label{sec:extremal}
Often the most urgent question is \emph{when the first member arrives}; unlike a programmed functional, this \emph{extremal functional} depends only on the support of $\mu$, not on the members' masses. For a population outside a boundary $R_b$, identify $\xi$ with $(R,v)$: at constant speed, a member arrives after $t_a(R,v)=(R-R_b)/v$ if $v>0$ and never ($t_a=+\infty$) otherwise, and $t_{\min}(\mu)=\inf\{t_a(\xi):\xi\in\supp\mu\}$. On $\{R>R_b,\,v>0\}$, $t_a$ increases in $R$ and thus has no local maximum, so every point of a level set of $t_a$ is a limit of points with later arrival times. With finitely many continuous kernels, later members can therefore mimic a first arrival in every readout.
\begin{theorem}[Finite probe sets cannot determine extremal functionals]\label{thm:impossible}
Under Assumption~\ref{as:gaussian}, let $t_a:\Xi\to\R$ on a domain $\Xi\subset\R^2$ have a nonempty level set $\{t_a=t_0\}$ free of local maxima, and let $g_1,\dots,g_J$ be the kernels of a finite set of probes plus the real and imaginary parts of the cross-kernels $\chi_{s,h_i}\chi_{s,h_j}^*$ of templates reading a common return. Then there are finite measures $\mu,\mu'$ on $\Xi$ with $\int g_i\,d\mu=\int g_i\,d\mu'$ for all $i$, hence identical readout laws, also under adaptive choice among them, and $t_{\min}(\mu)=t_0<t_{\min}(\mu')$, so no estimator, even from noiseless readouts, determines $t_{\min}$.
\end{theorem}
If the vectors $(g_i(\xi))_i$ on $\Xi_1=\{t_a\ge t_1\}$, $t_1>t_0$, span those of $\Xi$, the construction keeps $\mu'$ in $\Xi_1$, so the readouts cannot even tell $t_{\min}=t_0$ from $t_{\min}\ge t_1$; real-analytic kernels, such as the sweep's, satisfy this if $\Xi_1$ has an interior point. Nor does a minimum member mass help, since scaling $\mu$ and $\mu'$ preserves their equal readouts. A \emph{mass margin}, a first-arrival mass above the leakage of later arrivals into the readout ($\overline m<\underline m$ below), stops them from mimicking the first.
\begin{theorem}[Adaptive bisection]\label{thm:bisection}
Under Assumption~\ref{as:gaussian}, let $t_{\min}\in[t_{\mathrm{lo}},t_{\mathrm{hi}}]$, and for each horizon $t$ let a cumulative readout $\hat F_t$ over $n_t$ orthonormal templates have mean $m_t\le\overline m$ if $t_{\min}\ge t+\delta_r$ and $m_t\ge\underline m>\overline m$ if $t_{\min}\le t-\delta_r$, for a resolution $\delta_r>0$. Bisect, keeping the earlier half if $\hat F_t>\Theta=(\overline m+\underline m)/2$. For any $\delta,p_{\mathrm e}>0$, with $M\ge4(\underline m+n_t\sigma^2)^2/((\underline m-\overline m)^2p_{\mathrm e})$ looks per test, each decision errs with probability at most $p_{\mathrm e}$, and after $L=\lceil\log_2\frac{t_{\mathrm{hi}}-t_{\mathrm{lo}}}{\delta}\rceil$ tests the interval has width at most $\delta$ and contains $t_{\min}$ up to $\delta_r$ with probability at least $1-Lp_{\mathrm e}$.
\end{theorem}
For the chirp, the test at horizon $t$ is the sweep with $\alpha=\cc/(\lambda t)$ summed over $\mathcal B$ (Proposition~\ref{prop:cumulative}); $\underline m$ is the smallest readout mass of a member times the fraction left after bin straddling, $\overline m$ the largest total sidelobe leakage of later arrivals, and $\delta_r$ the largest range bin divided by the slowest closing speed. With $L$ tests, a nonadaptive bank leaves a worst-case interval of at least $(t_{\mathrm{hi}}-t_{\mathrm{lo}})/(L+1)$, and bisection one of $(t_{\mathrm{hi}}-t_{\mathrm{lo}})/2^L$ with $LM=O(L^2/P_{\mathrm e})$ looks for overall error probability $P_{\mathrm e}=Lp_{\mathrm e}$. Because members move between tests, the tests refer $t_{\min}$ to one instant: for constant speeds every arrival time decreases by the elapsed time, so a test taken $\Delta t$ after that instant checks $t_{\min}\le t$ with horizon $t-\Delta t$, and its duration adds to $\delta_r$.

\section{Sensing Procedure and Numerical Results}\label{sec:numerical}
\textbf{Procedure.} The task fixes a boundary $R_b$ and a bank of horizons $\Tst_1<\dots<\Tst_H$ (negative for down-chirps) at which masses, fluxes, or moments are wanted. For horizon $l$, the sensor transmits $M$ chirps (MTI pairs under clutter or transmit leakage) of rate $\alpha_l=\cc/(\lambda\Tst_l)$ and duration $T_l=\min\{\sqrt{\lambda|\Tst_l|/(\gamma v_{\max})},B_{\max}/|\alpha_l|\}$, the stop-and-hop limit of Assumption~\ref{as:stophop} capped at a bandwidth $B_{\max}$; below the cap, the range resolution is $\sqrt{\gamma\lambda v_{\max}|\Tst_l|}/2$ ($1.08$~m at $3$~s). Each return is dechirped, windowed (Hann for moments), and transformed by a DFT, whose bins collect the members predicted to share a range $\Tst_l$ ahead. Subtracting the noise floor $\sigma^2$, estimated from bins beyond the members' reach, and averaging over looks gives the profile estimate $\hat p_l(b)=\sum_i(|z_{b,i}|^2-\sigma^2)/(MT_l)$. Its sums over the bins inside $R_b$ give the masses and fluxes of Section~\ref{sec:chirp}, its rescaled bins the counts, least-squares fits of its centroids and spreads to \eqref{eq:moments} give $\hat{\bar v}$ and $\hat V_v$, and bisection over horizons reads the first arrival.

\textbf{Practical considerations.} The chirps of all horizons are interleaved to share one mean time, so motion during a frame of duration $T_f$ shifts every centroid equally, adding no mean-speed bias, and smears the profiles by at most $v_{\max}T_f$ ($1.3$~m for four looks of the $1$--$3$~s bank). An offset local oscillator lifts the slow sweeps' beat band (Table~\ref{tab:resources}) above the intermediate-frequency high-pass; transmit leakage and direct-current offsets, at zero beat frequency, cancel in the MTI pair (Proposition~\ref{prop:mti}); and range correlation suppresses oscillator phase noise by about $(2\pi f\tau_k)^2$ at offset $f$ ($-44$~dB at $1$~kHz for $\tau_k=1$~\textmu s).

\textbf{Setup.} The experiments simulate the beat signal \eqref{eq:beat}, including $\kappa_k$ (at $v_{\max}$, a $5.3\%$ loss at $f_k$), at 77~GHz with $v_{\max}=25$~m/s and $\gamma=16$; the horizons $\Tst\in\{1,2,3\}$~s give $(T,B)=(3.1\,\mathrm{ms},240\,\mathrm{MHz})$, $(4.4\,\mathrm{ms},170\,\mathrm{MHz})$, and $(5.4\,\mathrm{ms},139\,\mathrm{MHz})$. Unless stated otherwise, $K=300$ members of unit reflectivity have ranges uniform on $[60,150]$~m and closing speeds uniform on $[5,25]$~m/s, amplitudes follow Assumption~\ref{as:gaussian}, $R_b=40$~m, and the member SNR is $\rho E/\sigma^2$. The baselines apply a two-dimensional DFT, with range--Doppler coupling compensated, to a CPI of $36$~\textmu s chirps with the same dwell, bandwidth, and energy, followed by map integration or by peak detection at false-alarm rate $10^{-4}$ on a Hann-windowed map and counting. All estimators assume only that members lie beyond $R_b$ and close at most at $v_{\max}$: the bin sum, a programmed readout, adds the $72$ bins (at $3$~s) of predicted range in $[R_b-v_{\max}\Tst-2|\Delta_R|,R_b]$, and map integration sums the cells of these predicted ranges, at any speed, whose current range is at least $R_b-2|\Delta_R|$, $40$ per bin on average. The floor is known unless stated. Monte Carlo (MC) points use $40$--$200$ trials ($\pm$: one standard error), and exact values from Theorem~\ref{thm:variance} include the bias.

\begin{figure}[t]\centering\includegraphics[width=\columnwidth]{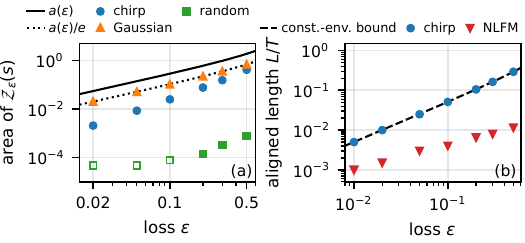}
\caption{Alignment geometry. (a) Area of $\mathcal Z_\varepsilon(s)$ against $a(\varepsilon)$ of Theorem~\ref{thm:geometry} (dotted: $a(\varepsilon)/e$) for a chirp ($TB=200$), a Gaussian pulse, and a random-phase waveform (hollow: grid-limited). (b) Longest pairwise-aligned segment of a line of slope $\alpha$ against the constant-envelope bound of Theorem~\ref{thm:line}(ii) (NLFM: nonlinear frequency modulation).}\label{fig:alignment}\end{figure}
\begin{figure*}[t]\centering\includegraphics[width=\textwidth]{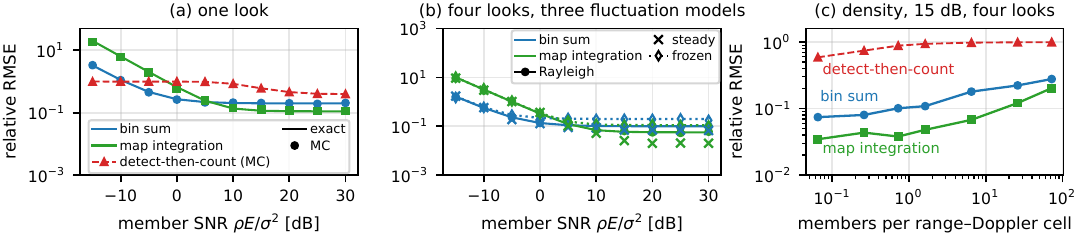}
\caption{Relative root-mean-square error (RMSE) of the mass inside $R_b$ at $\Tst=3$~s. (a) Against member SNR, one look (exact: Theorem~\ref{thm:variance}). (b) Four looks, block Rayleigh (Assumption~\ref{as:gaussian}), steady, and frozen amplitudes (solid: exact for Rayleigh; dotted: exact for frozen). (c) MC against density, $15$~dB, four looks.}\label{fig:snr}\end{figure*}
\begin{figure*}[t]\centering\includegraphics[width=\textwidth]{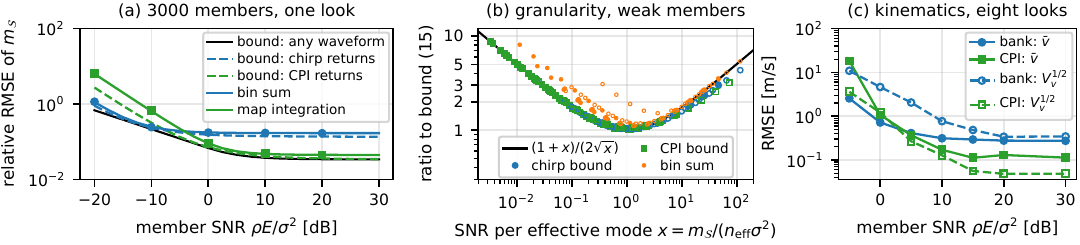}
\caption{Limits, granularity ($\Tst=3$~s), and kinematics. (a) Exact RMSEs (markers: MC) and square roots of bounds \eqref{eq:limit} (chirp's and CPI's returns, known positions) and \eqref{eq:universal}, relative to $m_\sS$. (b) Bounds \eqref{eq:limit} and bin-sum RMSE (at the chirp's $x$) over \eqref{eq:universal}, all as in (a), against the SNR per mode; uniform (filled) and clustered (hollow) scenes, $-25$ to $0$~dB. (c) MC RMSE of $\bar v$ and $V_v^{1/2}$: chirp bank (Proposition~\ref{prop:moments}) and CPI map of equal dwell and energy.}\label{fig:limits}\end{figure*}

\textbf{Alignment (Fig.~\ref{fig:alignment}).} On a fine grid, the area of $\mathcal Z_\varepsilon(s)$ lies below $a(\varepsilon)$ for all three waveforms and, for the Gaussian, which attains \eqref{eq:lieb}, matches $a(\varepsilon)/e=-\ln(1-\varepsilon)$ within about $0.5\%$: the bound is tight up to the factor $e$. The chirp's region is a needle of area of order $\varepsilon^{3/2}$. For $24$ unit members, the best template on $[0,T)$ captures a fraction $0.85$ of their mass (Theorem~\ref{thm:gram}) along a $0.23T$ segment of the chirp's own line, $0.21$ in an $8\times8$-cell blob, and $0.045\approx1/24$, one member's worth, on a line of opposite slope. On lines, the chirp attains the bound of Theorem~\ref{thm:line}(ii) at every loss, whereas a constant-envelope nonlinear frequency modulation (NLFM) chirp of the same mean slope (rate varying $\pm75\%$) aligns at most $0.012T$ and a random-phase waveform no segment.

\textbf{Predicted profiles.} For $120$ members of reflectivities $0.3$--$2$ at $10$~dB and $\Tst=2$~s, the expected profile of the exact dechirped echo (time-varying delay, all terms in $\beta_k$, onset) differs by at most $0.3\%$ of the peak from \eqref{eq:profile}; $400$-look means match it within $3.5\%$ (median), against their $5\%$ relative standard deviation. In the default scene at $10$~dB (Fig.~\ref{fig:profiles}(b)), one sweep per horizon follows the predicted range profile up to speckle. In occupied $4$~m bands, its root-mean-square error (RMSE) is $4.3$--$6.6$ members over $\Tst=1$--$3$~s, above the $3.0$--$4.2$ of one map with $66$--$77$ times its samples, whose resolved cells diversify the speckle; four sweeps reach $2.9$--$3.2$, still with $16$--$19$ times fewer samples. In empty bands, the sweep's errors are $0.3$--$0.4$ and the map's $1.2$--$2.6$, the noise of its many cells.

\textbf{Regime boundary (Fig.~\ref{fig:snr}(a)).} For both readout sums, the MC relative RMSE of the mass inside $R_b$ ($85$ members) matches Theorem~\ref{thm:variance} over $45$~dB of SNR. At $0$~dB and one look, map integration, which sums the noise of $40$ cells per bin, has an exact $0.66$ against the bin sum's $0.27$, a gap that widens as the SNR falls; at high SNR, the map's speckle diversity makes its RMSE about $1.8$ times smaller. The two cross at $5.8$~dB, $1.5$~dB below the $7.3$~dB of Corollary~\ref{cor:regime} for $n=40$ evenly filled cells per bin of mass $85/72$, because uneven bin filling doubles the bin sum's speckle excess over the map. Restricting the map to the true speed band ($34$ cells per bin) lowers the crossover by $0.4$~dB, as the $\sqrt n$ law predicts. Detect-then-count misses nearly all up to $5$~dB and still undercounts by $40\%$ at $30$~dB, as its $4\times4$-cell Hann main lobe, holding about one member at $0.065$ members per cell, merges neighbors.

\textbf{Fluctuation and density (Fig.~\ref{fig:snr}(b), (c)).} With four looks, steady scatterers (constant modulus, phases advancing with their own Doppler) and frozen amplitudes (no renewal) follow the block Rayleigh curves at low SNR, where the noise term dominates. At high SNR, frozen amplitudes forfeit the averaging of looks and lower the crossover by about $3$~dB, while steady ones lower the floors through their negative fourth-moment term (Section~\ref{sec:statistics}). Density biases only detection: at $15$~dB and four looks, compressing the population into smaller boxes around predicted range $R_b-20$~m (estimators now assuming only $R\ge0$) raises the detector's undercount from $59\%$ to $99\%$ at $72$ members per cell, near the bound $1-n/K$ of Section~\ref{sec:resources}, whereas the bin sum and map integration, both readout sums, stay nearly unbiased, their RMSE growing only as fewer modes are occupied (Theorem~\ref{thm:variance}).

\textbf{Estimation limits (Fig.~\ref{fig:limits}(a)).} A denser scene of $3000$ unit members ($0.65$ per cell) tests Corollary~\ref{cor:every}; the set $\sS$ inside $R_b$ after $3$~s holds $860$. With one look at $-10$~dB, the bin sum estimates $m_\sS$ with relative RMSE $0.250$ (MC $0.243\pm0.012$), including a $2.8\%$ bias from bins straddling $R_b$. This is below $0.319$, the bound \eqref{eq:limit} for every unbiased CPI estimator, even one that knows the positions, and only $16\%$ above $0.216$, the bound \eqref{eq:universal} for any waveform. The number of modes explains both comparisons. The chirp splits the mass of $\sS$ into $n_{\mathrm{eff}}=38$ modes in $50$ occupied bins, near unit SNR on average ($x=2.25$), whereas the CPI splits it into $528$ weak modes ($x=0.163$), each paying a noise term. The cosh law predicts bounds \eqref{eq:limit} $8\%$ and $44\%$ above \eqref{eq:universal} (actual: $9\%$ and $48\%$), and condition \eqref{eq:every} holds ($461<528$). Neither biased CPI estimator tried closes the gap: optimal shrinkage of the unbiased ones, an oracle needing the true mass, leaves at least $0.304$, and map integration has $0.65$.

As the SNR grows, the advantage passes to resolution. The bin sum leads the CPI bound up to about $-8$~dB, where the two SNRs per mode ($3.6$ and $0.26$) are nearly reciprocal, which the cosh law penalizes equally, and map integration up to about $-4$~dB. Above $0$~dB, as Proposition~\ref{prop:granularity} predicts, the CPI bound stays within $12\%$ of \eqref{eq:universal} and map integration within $31\%$, while the bin sum levels off near $0.17$, the speckle of its $38$ modes.

\textbf{Optimal granularity (Fig.~\ref{fig:limits}(b)).} To test the cosh law of Section~\ref{sec:limits} beyond this scene, we vary density (uniform scenes of $0.065$--$1.3$ members per cell), clustering ($3000$ members in $10$--$1000$ clusters of $2$~m by $0.5$~m/s), and SNR ($-25$ to $0$~dB). Throughout, the chirp and CPI bounds over \eqref{eq:universal} follow $(1+x)/(2\sqrt x)$ within $8.4\%$ for $x\le10$ (and lie at most $26\%$ below it beyond, where modes are unequal), so the comparison is decided by how each waveform splits the mass into modes. With density, the chirp's mode count saturates ($24$ to $38$) while the CPI's grows ($84$ to $756$), so the chirp's $x$ rises (from $0.36$ to $4.6$ at $-10$~dB); at $1.3$ members per cell, $x=4.6$ and $0.23$ are nearly reciprocal, and the known-position CPI bound ties the bin sum ($0.202$ against $0.205$). Clustering reduces the CPI's modes and raises both SNRs per mode, the CPI's toward one: with $30$ clusters at $-10$~dB ($x=4.6$ and $2.3$), the CPI bound of $0.213$ falls below the bin sum's $0.256$. At $-25$~dB the bin sum beats the CPI bound in every scene but those with $30$ and $10$ clusters, whose CPI modes ($48$ and $9$), fewer than the $72$ summed bins, pay fewer noise terms, violating \eqref{eq:every}. At low $x$, these $72$ noise terms put the bin sum a factor $\sqrt{72/n_{\mathrm{eff}}}$ above the chirp bound.

\textbf{Robustness.} The bin sum tolerates noise-floor errors far better than map integration. With the floor estimated per look from $103$ empty bins of a $1.25$-times oversampled beat band, the bin sum's RMSE at $-10$~dB in Fig.~\ref{fig:limits}(a) rises only to $0.263$ (MC $0.279\pm0.015$), still below the CPI bound. A systematic floor error of $0.1$~dB biases the bin sum by $2\%$ but map integration by $79\%$, in proportion to their $72$ and $2911$ noise terms. Unequal member SNRs, from $R^{-4}$ path loss and $3$~dB lognormal cross sections ($5$--$95\%$ spread $13.5$~dB, mean $-10$~dB over $\sS$), barely change the bin sum ($0.242$, between $0.216$ for \eqref{eq:universal} and $0.282$ for the CPI bound). Accelerations test the linear prediction: independent ones, uniform in $\pm1$~m/s$^2$, change the true mass inside $R_b$ by only $2\%$, but a common deceleration $\dot v=-1$~m/s$^2$, which leaves every member $-\dot v(\Tst)^2/2=4.5$~m farther than predicted, changes it by $-15\%$, an error the bin sum inherits, staying within $7\%$ of the linear prediction.

\textbf{Clutter.} With $30$ static scatterers of reflectivity $3$ ($\Tst=2$~s, four looks, $10$~dB), a plain sweep reads a mass of $303$ in the predicted-range band $55$--$155$~m, where the movers hold $217$; the MTI template of Proposition~\ref{prop:mti} removes the clutter, reading $208$ ($220$ for the movers alone). For clutter with a Gaussian internal velocity spread $\sigma_v$, the weight $2\sin^2(\pi\nu T_p)$ averages to $1-\exp(-8\pi^2\sigma_v^2T_p^2/\lambda^2)$: the template ($T_p=4.4$~ms) suppresses it by $20$~dB at $\sigma_v=10$~mm/s and still by $6.5$~dB at $50$~mm/s, as MC confirms.

\textbf{Kinematic moments (Fig.~\ref{fig:limits}(c)).} With a Hann window and eight looks, a bank of four equal-power sweeps at $\Tst=\pm1$ and $\pm3$~s (Proposition~\ref{prop:moments}, $17.0$~ms per look) and CPI map moments of equal dwell and energy, with $56$ times more samples, estimate the mean closing speed of $15.13$~m/s; SNRs refer to the $5.4$~ms sweep. As for masses, the bank wins at $-5$ and $0$~dB and the map from $5$~dB up: its cells resolve the members, so its floor ($0.11$--$0.13$~m/s) is their speckle $V_v^{1/2}/\sqrt{KM}\approx0.12$~m/s, whereas the bank's bins aggregate members whose interference changes between horizons (floor $0.28$~m/s). For the speed spread $V_v^{1/2}$ ($5.66$~m/s), read from the curvature of $V(\Tst)$ in \eqref{eq:moments}, the map wins throughout.

\textbf{Counts.} Path loss and $3$~dB lognormal cross sections are redrawn per trial, with SNRs referred to a range $R_{\mathrm{ref}}$, over four looks. Each look transmits an up- and a down-chirp of the $3$~s design; their centroid difference gives $\hat{\bar v}$ (rectangular window, centroid bias at most $|\Delta_R|/(2\pi)$), and the up-chirp's bins are rescaled as in Section~\ref{sec:moments}; map integration, with $77$ times the samples, rescales every cell by its own range. In a flock of $300$ members at $110$--$150$~m closing at $15$~m/s (spread $1$~m/s), about $150$ of which are inside $R_b=85$~m after $3$~s ($R_{\mathrm{ref}}=130$~m), $\hat{\bar v}$ is accurate to $0.5$~m/s at $0$~dB and $0.2$~m/s from $10$~dB up, and the count's relative RMSE falls from $0.57$ at $-10$~dB to $0.12$--$0.16$ from $0$~dB up. Map integration ($0.91$ at $-10$~dB) matches it at $0$~dB and is better at high SNR ($0.08$--$0.09$) through speckle diversity, whereas detect-then-count misses $77$--$100\%$.

\textbf{First arrival.} For $40$ members ($20$~dB, eight looks, threshold $0.4$ member masses, horizons in $[0.5,8]$~s), adaptive bisection (Theorem~\ref{thm:bisection}) halves the interval containing $t_{\min}$ at every test, correct to $0.05$~s in $99$--$100\%$ of $200$ trials ($0.12$~s wide after six tests), while a nonadaptive bank of as many tests leaves $1.07$~s. On a moving scene, with six tests back to back over $0.19$~s, horizons referred to the start keep $99\%$ success, whereas tests ignoring elapsed time succeed in $12\%$.
\section{Discussion}\label{sec:discussion}
\textbf{Operating regimes.} At every SNR, the sweep keeps its per-horizon sample savings (Table~\ref{tab:resources}) and a floor-error bias smaller than map integration's by a factor of the map's cells per bin ($40$ here), whereas its accuracy gain for masses depends on the member SNR and density: the crossovers of Section~\ref{sec:numerical} lie between $-10$ and $6$~dB, falling with density, and strong clustering lowers or removes them. With $12$~dBm transmit power, $25$~dBi antenna gains, and a $12$~dB noise figure, one $5.4$~ms sweep at 77~GHz gives, at $150$~m and before system losses, a member SNR of about $13$~dB for a $-20$~dBsm bird (dBsm: dB re $1$~m$^2$), $3$~dB for a $-30$~dBsm small bird, and $-7$~dB for a $-40$~dBsm large insect. At this range, birds thus lie near or above the crossovers with map integration, where the sweep's gains are its samples and floor-error robustness, whereas insects at moderate densities lie below them, where the sweep is also more accurate.

\textbf{Extensions and open problems.} The theory applies to any sensor obeying the narrowband delay--Doppler model, such as coherent lidar and narrowband sonar, and a steered beam's two-way pattern, part of $\rho_k$, selects sector masses. In orthogonal frequency-division multiplexing (OFDM), a diagonal reference-signal pattern on the subcarrier--symbol grid forms a stepped-frequency chirp of programmable slope, so an ISAC base station could report predicted traffic profiles. Open problems include semiparametric bounds \cite{tsang2019semiparametric} without known positions; which spectra $|\Fs\Psi|$ are squared ambiguity surfaces (Theorem~\ref{thm:transfer}(iv)) \cite{sussman1962least} and which waveforms give unit-SNR modes (Proposition~\ref{prop:granularity}) at a prescribed horizon; NLFM for accelerating members; and wideband programs, with Doppler as time scaling, for ultrasound blood volume flow.

\section{Conclusion}
Over-the-scene computation turns illumination into a program and the scene into its computer, with the transmit ambiguity function as the transfer function. One readout aggregates members over only a small delay--Doppler area but along line segments, longest for the chirp among constant-envelope waveforms, so a chirp's mean beat spectrum is the predicted range profile. Because aggregation saves noise terms but forfeits speckle diversity, unbiased estimation ideally resolves bright members and groups weak ones into modes of unit SNR; in the low-SNR limit, a sum of bin readouts beats every unbiased estimator from an equal-energy CPI once it has fewer terms than the CPI has effective modes, and at every SNR it needs far fewer samples per horizon. First arrivals, which no finite probe set determines in unconstrained Rayleigh scenes, are localized by bisection given a mass margin.

\appendices
\section{Proofs for Section~\ref{sec:compute}}\label{app:compute}
\begin{lemma}\label{lem:weyl}
$\ip{\Tsh_\xi s}{\Tsh_{\xi'}s}=e^{j2\pi(\nu-\nu')\tau'}\chi_s(\xi-\xi')$ for $\xi=(\tau,\nu)$, $\xi'=(\tau',\nu')$, and $\Tsh_\zeta\Tsh_\xi=e^{j2\pi(\omega\tau-\nu x)}\Tsh_\xi\Tsh_\zeta$ for $\zeta=(x,\omega)$.
\end{lemma}
\begin{proof}
Substitute $u=t-\tau'$ in the inner product; for the commutation, $(\Tsh_\zeta\Tsh_\xi f)(t)=f(t-x-\tau)e^{j2\pi\nu(t-x)+j2\pi\omega t}$ and $(\Tsh_\xi\Tsh_\zeta f)(t)=f(t-\tau-x)e^{j2\pi\omega(t-\tau)+j2\pi\nu t}$.
\end{proof}

\begin{proof}[Proof of Theorem~\ref{thm:transfer}]
\emph{Spreading functions.} For a rank-one operator $R=ab^*$ put $\eta_R=\chi_{b,a}^*$. By \eqref{eq:moyal}, $\iint\eta_R(\zeta)\ip{\Tsh_\zeta f}{f'}d\zeta=\iint\chi_{f,f'}\chi_{b,a}^*=\ip{f}{b}\ip{a}{f'}=\ip{Rf}{f'}$ for all $f,f'$, and $\ip{\eta_{ab^*}}{\eta_{cd^*}}_{L^2}=\ip{a}{c}\ip{d}{b}=\ip{ab^*}{cd^*}_{\HS}$. By linearity, $\eta_Q$ represents a finite-rank $Q$, with $\norm{\eta_Q}_{L^2}=\norm{Q}_{\HS}$; $Q\mapsto\eta_Q$ extends to a unitary map from the Hilbert--Schmidt operators onto $L^2(\R^2)$ \cite{grochenig2001foundations}.

(i) Since $\Tsh_\xi$ is unitary, Lemma~\ref{lem:weyl} gives $\ip{\Tsh_\zeta\Tsh_\xi s}{\Tsh_\xi s}=e^{j2\pi(\omega\tau-\nu x)}\ip{\Tsh_\zeta s}{s}=e^{j2\pi(\omega\tau-\nu x)}\chi_s(\zeta)$. Representing $Q$ by $\eta_Q$ therefore gives
\begin{align}
g_{s,Q}(\xi)&=\iint\eta_Q(\zeta)\ip{\Tsh_\zeta\Tsh_\xi s}{\Tsh_\xi s}\,d\zeta\nonumber\\
&=\iint\eta_Q(\zeta)\chi_s(\zeta)\,e^{j2\pi(\omega\tau-\nu x)}\,d\zeta,\label{eq:pf-transfer}
\end{align}
which is $\Fs[\eta_Q\chi_s](\xi)$ with $\eta_Q\chi_s\in L^1\cap L^2$, as $\eta_Q,\chi_s\in L^2$ and $|\chi_s|\le E$ (Cauchy--Schwarz), and $\Fs^2=I$ gives (i).

(ii) \emph{Necessity.} Let $g_{s,Q_n}\to\Psi$ in $L^2$ with bounded $\norm{Q_n}_{\HS}$, and $\Lambda^2=\liminf_n\norm{Q_n}_{\HS}^2$. Along a subsequence with $\norm{Q_n}_{\HS}\to\Lambda$, a further subsequence of $\eta_{Q_n}$ converges weakly to some $\eta$ with $\norm{\eta}\le\Lambda$, and multiplication by the bounded $\chi_s$ is weakly continuous, so $\eta_{Q_n}\chi_s\rightharpoonup\eta\chi_s$; by (i) and the unitarity of $\Fs$, $\eta_{Q_n}\chi_s=\Fs g_{s,Q_n}\to\Fs\Psi$. Hence $\Fs\Psi=\eta\chi_s$, which vanishes where $\chi_s$ does, and $c_s(\Psi)=\iint_{\{\chi_s\ne0\}}|\eta|^2\le\Lambda^2$; $Q_n\equiv Q$ gives the last claim. \emph{Sufficiency.} If $c_s(\Psi)<\infty$, let $\eta=\Fs\Psi/\chi_s$ on $\{\chi_s\neq0\}$ and $\eta=0$ elsewhere, so that $\norm{\eta}^2=c_s(\Psi)$, and let $Q_\eta$ be the Hilbert--Schmidt operator with spreading function $\eta$. Since $|g_{s,Q}|\le E\norm{Q}_{\HS}$ for all $Q$ and $\norm{g_{s,Q}}_{L^2}=\norm{\eta_Q\chi_s}_{L^2}\le E\norm{Q}_{\HS}$ for finite-rank $Q$, (i) extends by density to $Q_\eta$ and $g_{s,Q_\eta}=\Fs[\eta\chi_s]=\Psi$. $Q_\eta$ is Hermitian, as $\eta_{Q^*}(\zeta)=e^{-j2\pi\omega x}\eta_Q^*(-\zeta)$, $\chi_s(\zeta)=e^{j2\pi\omega x}\chi_s^*(-\zeta)$ (Lemma~\ref{lem:weyl}), and $(\Fs\Psi)(-\zeta)=(\Fs\Psi)^*(\zeta)$ for real $\Psi$. Its spectral truncations $Q_n$ are finite-rank and Hermitian, satisfy $\norm{Q_n}_{\HS}\le\norm{Q_\eta}_{\HS}$, and converge to $Q_\eta$ in that norm, so $g_{s,Q_n}\to g_{s,Q_\eta}$, a continuous version of $\Psi$, uniformly and in $L^2$.

(iii) Write $Q=\sum_{i,j}q_{ij}h_ih_j^*$ with orthonormal $h_i$, so that $\norm{Q}_{\HS}=\norm{q}_F$, and let $\mathbf z=(\ip{r}{h_i})_i$. Under Assumption~\ref{as:gaussian}, $\mathbf z\sim\mathcal{CN}(\mathbf 0,\Sigma)$ with $\Sigma=A+\sigma^2I$, $A\succeq0$, and $\tr(qA)=\int g_{s,Q}\,d\mu$, so $\hat F_Q$ is unbiased, and the variance $\tr(q\Sigma q\Sigma)$ of $\mathbf z^{\mathsf H}q\mathbf z$ splits into three nonnegative terms:
\begin{align}
M\Var\hat F_Q&=\tr\big((A^{\frac12}qA^{\frac12})^2\big)+2\sigma^2\tr(A^{\frac12}q^2A^{\frac12})+\sigma^4\norm{q}_F^2\nonumber\\
&\ge\sigma^4\norm{Q}_{\HS}^2 .\label{eq:pf-hs}
\end{align}
(iv) If $c_s(\Psi)<\infty$, Cauchy--Schwarz and $\iint|\chi_s|^2=E^2$ give $\norm{\Fs\Psi}_{L^1}^2=(\iint_{\{\chi_s\ne0\}}\frac{|\Fs\Psi|}{|\chi_s|}|\chi_s|)^2\le c_s(\Psi)E^2$, with equality iff $|\Fs\Psi|/|\chi_s|\propto|\chi_s|$.
\end{proof}

\begin{proof}[Proof of Theorem~\ref{thm:gram}]
Here $\ip{\Gamma_sh}{h}=\sum_k\rho_k|\ip{u_k}{h}|^2$ equals $\int g_{s,h}\,d\mu$, and since $\Gamma_s\succeq0$, $\ip{\Gamma_sh}{h}\le\vartheta_{\max}\le\tr\Gamma_s=E\mu(\R^2)$, with equality on the right iff $\rank\Gamma_s=1$. (i) With $h=u_{k_0}/\sqrt E$, $\ip{\Gamma_sh}{h}\ge\sum_{k\in\sS}\rho_k|\ip{u_k}{u_{k_0}}|^2/E\ge(1-\varepsilon)E\mu(\sS)$. (ii) Weight Bessel's inequality $\sum_{k\in\sS}|\ip{u_k}{h}|^2\le E$ by $\rho_k\le\max_{k\in\sS}\rho_k$.
\end{proof}

\begin{proof}[Proof of Theorem~\ref{thm:geometry}]
(i) On $\mathcal Z_\varepsilon(s)$, $|\chi_s|^p\ge(1-\varepsilon)^{p/2}E^p$, so \eqref{eq:lieb} gives $|\mathcal Z_\varepsilon(s)|\le(2/p)(1-\varepsilon)^{-p/2}$ for every $p\ge2$. With $\ell=\ln\frac1{1-\varepsilon}$, $(2/p)e^{p\ell/2}$ is minimized at $p=2/\ell$, with value $e\ell$, if $\ell\le1$, and at $p=2$ otherwise. (ii) $\mathcal A-\mathcal A\subset\mathcal Z_\varepsilon(s)$ by definition, and the Brunn--Minkowski inequality \cite{gardner2002brunn} gives $|\mathcal A-\mathcal A|\ge(|\mathcal A|^{1/2}+|-\mathcal A|^{1/2})^2=4|\mathcal A|$. (iii) On the overlap, of length $T-|\tau|$, $s(t-\tau)s^*(t)=e^{j\pi\alpha\tau^2}e^{-j2\pi\alpha\tau t}$, and $E=T$.
\end{proof}

\begin{proof}[Proof of Corollary~\ref{cor:tradeoff}]
With $a=\Tsh_\xi s/\sqrt E$, $b=\Tsh_{\xi'}s/\sqrt E$, and $g=g_{s,h}$, $g(\xi)-g(\xi')=E\ip{(aa^*-bb^*)h}{h}$ for unit-norm $h$, and $aa^*-bb^*$ has rank at most two, trace zero, and $\tr[(aa^*-bb^*)^2]=2-2|\ip ab|^2$, so its eigenvalues are $0$ and $\pm(1-|\ip ab|^2)^{1/2}$, and $|g(\xi)-g(\xi')|\le E\sqrt\varepsilon$ as $|\ip ab|^2=|\chi_s(\xi-\xi')|^2/E^2\ge1-\varepsilon$ (Lemma~\ref{lem:weyl}). With $m=\mu_1(\mathcal A)$, $\int g\,d\mu_1-\int g\,d\mu_2=m^{-1}\iint(g(\xi)-g(\xi'))\,d\mu_1(\xi)\,d\mu_2(\xi')$, of modulus at most $Em\sqrt\varepsilon$.
\end{proof}

\begin{proof}[Proof of Theorem~\ref{thm:line}]
(i) With $s=s_\alpha e^{j\pi\alpha t^2}$, the quadratic phases in $\int s(t-\Delta)e^{j2\pi\alpha\Delta t}s^*(t)dt$ combine to $\pi\alpha\Delta^2$; points of a line of slope $\alpha$ differ by $(\Delta,\alpha\Delta)$, and Lemma~\ref{lem:weyl} applies. (ii) The integrand $s_\alpha(t-\Delta)s_\alpha^*(t)$ has modulus $E/T$ on an overlap of length $T-|\Delta|$, so $|\chi_s(\Delta,\alpha\Delta)|\le E\max\{1-|\Delta|/T,0\}$, with equality iff its phase is constant on the overlap, as it is for the chirp. Conversely, write $s_\alpha=\sqrt{E/T}\,e^{j\phi}$ with $\phi$ continuous. Equality at lag $\Delta$ makes $\phi(t)-\phi(t-\Delta)$ continuous in $t$ on the overlap with values in $\Theta(\Delta)+2\pi\mathbb Z$, hence equal to a constant $\Theta(\Delta)$; evaluated at a point common to all overlaps with $\Delta<\Delta_0$ (taking $\Delta_0<T$), $\Theta$ is continuous. Telescoping, $\phi(t)-\phi(t-\Delta_1-\Delta_2)=\Theta(\Delta_1)+\Theta(\Delta_2)$, so $\Theta$ is additive on $(0,\Delta_0)$ and therefore linear, $\Theta(\Delta)=\omega_0\Delta$; then $\phi(t)-\omega_0t$ is invariant under all small shifts, hence constant, and $s$ is the chirp of rate $\alpha$ with $f_0=\omega_0/(2\pi)$.
\end{proof}

\section{Proofs for Sections~\ref{sec:chirp}--\ref{sec:extremal}}\label{app:chirp}
\begin{proof}[Derivation of \eqref{eq:beat} and proof of Theorem~\ref{thm:profile}]
For member $k$, whose echo is $a_ks(t-\tau_k(t))$, the product $s\,r^*$ has phase $2\pi[f_c\tau_k(t)+\alpha t\tau_k(t)-\alpha\tau_k(t)^2/2]$ with $\tau_k(t)=\tau_k-\beta_kt$, i.e., linear coefficient $2\pi[\alpha\tau_k(1+\beta_k)-f_c\beta_k]$, and quadratic coefficient $-2\pi\alpha\beta_k(1+\beta_k/2)$; with $\nu_k=f_c\beta_k$ and relative corrections of order $\beta_k$ dropped, this is \eqref{eq:beat}, and the beat noise is white because $|s|=1$. Proposition~\ref{prop:identity}, applied to the responses $e^{j\theta_k}e^{j2\pi f_kt-j2\pi\kappa_kt^2}$ on $[\tau_k,T)$, gives \eqref{eq:profile}. Extending the integral to $[0,T)$ changes $W_k$ by at most $\norm{w}_\infty\tau_k$, and $|e^{-j2\pi\kappa_kt^2}-1|\le2\pi|\kappa_k|t^2$ gives $\norm{w}_\infty\int_0^T2\pi|\kappa_k|t^2dt\le\frac{4\pi}{3\gamma}\norm{w}_\infty T$ by Assumption~\ref{as:stophop}. By \eqref{eq:beat}, $f_k=f_b$ is the locus \eqref{eq:locus}, and a beat-frequency width $1/T$ is $\cc/(2B)$ in $R_0$.
\end{proof}
\begin{proof}[Proof of Proposition~\ref{prop:mti}]
Over the pair, member $k$, of unit-amplitude beat signal $\varphi_k(t)=e^{j2\pi f_kt-j2\pi\kappa_kt^2}$, contributes $a_k^*e^{j\theta_k}\varphi_k$ and $a_k^*e^{j\theta_k}e^{-j2\pi\nu_kT_p}\varphi_k$ (up to the beat-frequency shift $2\alpha v_kT_p/\cc$, at most $2/\gamma$ of a bin for $T_p=T$), so the normalized difference has response $(1-e^{-j2\pi\nu_kT_p})\varphi_k/\sqrt2$, scaling every kernel by $2\sin^2(\pi\nu_kT_p)$, with noise density $\sigma^2$.
\end{proof}
\begin{proof}[Proof of Proposition~\ref{prop:moments}]
By Theorem~\ref{thm:profile}, for $f_b=b/T$, $p_{\Tst}(b)=\sum_k\rho_k\Omega(b-x_k)$, where $x_k=f_kT$ and $\Omega(y)=|\widehat W(y/T)|^2/T$ has Fourier transform $\widehat\Omega(u)=R_w(Tu)$, $R_w(\tau)=\int w(t)w(t+\tau)dt$. For the Hann window $|\widehat W(f)|^2=O(|f|^{-6})$, so $y^q\Omega(y)$, $q\in\{0,1,2\}$, is integrable with transform $(jT/2\pi)^qR_w^{(q)}(Tu)$. Poisson's summation formula then gives $\sum_{b\in\mathbb Z}(b-x)^q\Omega(b-x)=(\frac{jT}{2\pi})^q\sum_{l\in\mathbb Z}R_w^{(q)}(lT)\,e^{-j2\pi lx}$ for every $x$. Since $w$ vanishes to second order at $0$ and $T$, $R_w(\tau)=O((T-|\tau|)^5)$ near $|\tau|=T$ and $R_w=0$ beyond, so only $l=0$ remains, and the three sums are $R_w(0)=1$ (for $q=0$ and any $w$ on $[0,T]$, the unit sum in Theorem~\ref{thm:profile}), $R_w'(0)=0$, and $-T^2R_w''(0)/(4\pi^2)=\frac{T^2}{4\pi^2}\int|w'|^2=\frac13$, independent of $x$. With $R_0(b)=b\Delta_R$ and $R_k-v_k\Tst=x_k\Delta_R$, $\Delta_R=\cc/(2\alpha T)$, expanding $b=(b-x_k)+x_k$ in the first two moments of $p_{\Tst}$ and normalizing by $\mu(\R^2)$ gives \eqref{eq:moments}. For the rectangular window, $(b-x)\sinc^2(b-x)=\sin^2(\pi x)/(\pi^2(b-x))$ decays only as $1/|b|$; summed symmetrically about $x$, the partial-fraction expansion of $\pi\cot\pi x$ gives $\frac{\sin^2\pi x}{\pi^2}\sum_b\frac1{b-x}=-\frac{\sin2\pi x}{2\pi}$, and $\sum_b(b-x)^2\sinc^2(b-x)$ diverges for $x\notin\mathbb Z$.
\end{proof}
\begin{proof}[Proof of Theorem~\ref{thm:variance}]
Per look, $\mathbf z=W\mathbf a+\mathbf n\sim\mathcal{CN}(\mathbf 0,\Sigma)$ with $\mathbf a=(a_k)_k$ and $\mathbf n$ the noise readouts, $\Sigma=A+\sigma^2I$, $A=WDW^{\mathsf H}$. By the complex Isserlis formula $\Cov(|z_b|^2,|z_{b'}|^2)=|\Sigma_{bb'}|^2$, so $\Var\sum_b|z_b|^2=\tr(\Sigma^2)=\norm{G}_F^2+2\sigma^2\tr G+n\sigma^4$. The rank inequalities follow from Cauchy--Schwarz and the nonnegativity of the at most $\min(n,K)$ nonzero eigenvalues of $G\succeq0$.
\end{proof}
\begin{proof}[Proof of Theorem~\ref{thm:limits} and Proposition~\ref{prop:granularity}]
\emph{Reduction.} Let $\Phi$ be an orthonormal basis of the span of the responses and templates and $\mathbf y=\Phi^*r$, so that every readout, hence $\hat m$, is a function of the $M$ independent looks of $\mathbf y\sim\mathcal{CN}(\mathbf 0,\Sigma)$, $\Sigma=\Phi^*\Gamma_s\Phi+\sigma^2I$, $\tilde u_k=\Phi^*u_k$, whose density $p$ has Fisher information $\mathcal J_{kl}=M\tr(\Sigma^{-1}\partial_k\Sigma\,\Sigma^{-1}\partial_l\Sigma)=M|\tilde u_k^{\mathsf H}\Sigma^{-1}\tilde u_l|^2$, $\partial_k=\partial/\partial\rho_k$ \cite{kay1993fundamentals}.

\emph{A Cram\'er--Rao bound along one direction.} Differentiating $\E\hat m=E\sum_{k\in\sS}\rho_k$ along $\mathbf t$ gives $\Cov(\hat m,\mathbf t^{\mathsf T}\nabla\log p)=E\mathbf t^{\mathsf T}\mathbf 1_\sS$, with $\mathbf 1_\sS$ the indicator vector of $\sS$, so $\Var\hat m\ge E^2(\mathbf t^{\mathsf T}\mathbf 1_\sS)^2/\mathbf t^{\mathsf T}\mathcal J\mathbf t$ for every $\mathbf t$. The direction $t_k=\rho_k$ on $\sS$ and $0$ elsewhere gives, with $U=[\sqrt{\rho_k}\tilde u_k]_{k\in\sS}$,
\begin{equation}\label{eq:pf-crb}
\Var\hat m\ge\frac{m_\sS^2}{M\,\norm{U^{\mathsf H}\Sigma^{-1}U}_F^2}.
\end{equation}

\emph{Bound \eqref{eq:limit}.} With $P=U^{\mathsf H}U$, whose nonzero eigenvalues are the $\vartheta_i$, $\Sigma\succeq UU^{\mathsf H}+\sigma^2I$ and the push-through identity give $0\preceq U^{\mathsf H}\Sigma^{-1}U\preceq U^{\mathsf H}(UU^{\mathsf H}+\sigma^2I)^{-1}U=P(P+\sigma^2I)^{-1}$. Eigenvalues are monotone in Loewner order, so $\norm{U^{\mathsf H}\Sigma^{-1}U}_F^2\le\sum_i\psi(\vartheta_i)$, i.e., \eqref{eq:limit}; $\psi\le1$ and $\psi(\vartheta)\le\vartheta^2/\sigma^4$, with $\sum_i\vartheta_i^2=m_\sS^2/n_{\mathrm{eff}}$, give its floors.

\emph{Bound \eqref{eq:universal}.} $P$ is an $|\sS|\times|\sS|$ positive semidefinite matrix with diagonal entries $\rho_k\norm{u_k}^2=m_k$, so by Schur's theorem its eigenvalues, the $\vartheta_i$ padded with zeros, majorize $(m_k)_{k\in\sS}$ \cite{marshall2011inequalities}. The function $\hat\psi$ is concave: it is linear up to $\sigma^2$, where it meets $\psi$ with the common slope $1/(4\sigma^2)$, and $\psi''<0$ on $(\sigma^2/2,\infty)$. Since $\psi\le\hat\psi$ and $\hat\psi(0)=0$, Karamata's inequality \cite{marshall2011inequalities} yields $\sum_i\psi(\vartheta_i)\le\sum_i\hat\psi(\vartheta_i)\le\sum_{k\in\sS}\hat\psi(m_k)$. The right side depends on the waveform only through its energy, so it bounds each look's share of $\mathbf t^{\mathsf T}\mathcal J\mathbf t$ also when the waveform changes between looks; with \eqref{eq:pf-crb} this is \eqref{eq:universal}, and $\hat\psi(\vartheta)\le\min\{1,\vartheta/(4\sigma^2)\}$ gives its floors. Equality in both steps forces nonzero $\vartheta_i\ge\sigma^2$ (as $\psi<\hat\psi$ on $(0,\sigma^2)$) and tight Jensen steps $\hat\psi(m_k)\ge\sum_i|V_{ki}|^2\hat\psi(\vartheta_i)$, with $m_k=\sum_i|V_{ki}|^2\vartheta_i$ for unitary eigenvectors $V$ of $P$; as $\hat\psi$ is linear up to $\sigma^2$ and strictly concave beyond, the $\vartheta_i$ with $V_{ki}\ne0$ then equal $m_k$ if $m_k>\sigma^2$, so $P_{lk}=0$ for $l\ne k$, and lie in $\{0,\sigma^2\}$ otherwise.

\emph{Efficiency.} With $\langle X,Y\rangle_\Sigma=\tr(\Sigma^{-1}X\Sigma^{-1}Y)$, $\mathcal J_{kl}=M\langle\partial_k\Sigma,\partial_l\Sigma\rangle_\Sigma$. For $Z=\Sigma Q\Sigma$, the target $\tr(Q(\Sigma-\sigma^2I))$ has gradient $c_k=\tr(Q\partial_k\Sigma)=\langle Z,\partial_k\Sigma\rangle_\Sigma$. The projection $\Pi Z=\sum_ld_l\partial_l\Sigma$ of $Z$ onto $\operatorname{span}\{\partial_k\Sigma\}$ satisfies $(\mathcal J/M)\mathbf d=\mathbf c$, so the Cram\'er--Rao bound is $\mathbf c^{\mathsf T}\mathcal J^\dagger\mathbf c=\norm{\Pi Z}_\Sigma^2/M$, while $\Var\hat F_Q=\tr(Q\Sigma Q\Sigma)/M=\norm{Z}_\Sigma^2/M$; equality holds iff $Z=\Pi Z$. In the grouped case, $u_k=e^{j\phi_k}\sqrt E\,e_\ell$ for member $k$ of group $\ell$ with orthonormal $e_\ell$, $Q=\sum_\ell e_\ell e_\ell^*$, and $\Sigma Q\Sigma=\sum_\ell(m_\ell+\sigma^2)^2e_\ell e_\ell^*$ lies in the span; the target is $m_\sS$ and the variance $\sum_\ell(m_\ell+\sigma^2)^2/M$. For equal masses, $\sum_k\hat\psi(m_k)=|\sS|\hat\psi(m_k)$ gives the two values of \eqref{eq:universal}, which the grouped variance attains: $|\sS|(m_k+\sigma^2)^2/M$ for groups of one and $(m_\sS/\sigma^2)(2\sigma^2)^2/M$ for groups of mass $\sigma^2$.
\end{proof}
\begin{proof}[Proof of Theorem~\ref{thm:impossible}]
Let $\mathbf g(\xi)\in\R^J$ collect the continuous $g_i$, $S=\{t_a>t_0\}$, and $\mathcal V=\operatorname{span}\{\mathbf g(\xi):\xi\in S\}$, closed in $\R^J$. A point $\xi_0$ with $t_a(\xi_0)=t_0$, not a local maximum, is a limit of points of $S$, so $\mathbf g(\xi_0)=\sum_j\varsigma_j\mathbf g(\xi_j)\in\mathcal V$ with $\xi_j\in S$. For any $\xi_s\in S$, the measures $\mu=\delta_{\xi_0}+\delta_{\xi_s}+\sum_{\varsigma_j<0}|\varsigma_j|\delta_{\xi_j}$ and $\mu'=\delta_{\xi_s}+\sum_{\varsigma_j>0}\varsigma_j\delta_{\xi_j}$ have $\int\mathbf g\,d\mu=\int\mathbf g\,d\mu'$ but $t_{\min}(\mu)=t_0<t_{\min}(\mu')$. Under Assumption~\ref{as:gaussian} the readouts are zero-mean circular Gaussian, independent across looks, with covariances linear in $\mu$ via these kernels, so their laws under $\mu$ and $\mu'$ coincide, for adaptive probes too.
\end{proof}
\begin{proof}[Proof of Theorem~\ref{thm:bisection}]
By Theorem~\ref{thm:variance} and $\norm{G}_F^2\le(\tr G)^2$, $\Var\hat F_t\le(m_t+n_t\sigma^2)^2/M$. If $t_{\min}\ge t+\delta_r$, Chebyshev's inequality gives $\Pr\{\hat F_t>\Theta\}\le\frac{4(\overline m+n_t\sigma^2)^2}{M(\underline m-\overline m)^2}\le p_{\mathrm e}$; if $t_{\min}\le t-\delta_r$, $\Pr\{\hat F_t\le\Theta\}\le\frac{(m_t+n_t\sigma^2)^2}{M(m_t-\Theta)^2}$, which, as $(m+n_t\sigma^2)/(m-\Theta)$ decreases for $m>\Theta$, is at most its value at $m_t=\underline m$, itself at most $p_{\mathrm e}$. If no decision errs, $t_{\min}$ stays within $\delta_r$ of the current interval: a decision with $|t_{\min}-t|<\delta_r$ makes $t$ an endpoint, and any other keeps $t_{\min}$ on the retained side; a union bound over the $L$ decisions concludes.
\end{proof}

\end{document}